\documentclass[12pt]{article}
\usepackage[top=1in, bottom=1in, left=1.25in, right=1.25in]{geometry}
\usepackage{float}
\usepackage{color}
\usepackage{tikz}
\usepackage{setspace}
\usepackage[toc,page]{appendix}
\usepackage{amssymb}
\usepackage{showlabels}
\usepackage[normalem]{ulem}
\usepackage{multirow}
\usepackage{tikz}
\usepackage[toc]{appendix}
\usepackage{chngcntr}
\usepackage{authblk}
\usetikzlibrary{shapes,arrows}
\usepackage{subfigure}
\usepackage{cleveref}
\usepackage[round]{natbib}
\usepackage{latexsym}
\usepackage{amsmath}
\usepackage{graphicx}
\usepackage{caption}
\providecommand{\keywords}[1]{\textbf{\textit{Key words:}} #1}
\usepackage{bm}
\usepackage[pdftex,hypertexnames=false,linktocpage=true]{hyperref}
\hypersetup{colorlinks=true,linkcolor=blue,anchorcolor=blue,citecolor=blue,filecolor=blue,urlcolor=blue,bookmarksnumbered=true,pdfview=FitB}

\newcommand{\biblist}{\begin{list}{}
		{\listparindent 0.0cm \leftmargin 0.50cm \itemindent -0.50 cm
			\labelwidth 0 cm \labelsep 0.50 cm
			\usecounter{list}}\clubpanelty4000\widowpanelty4000}
	\newcommand{\ebiblist}{\end{list}}

\newtheorem{theorem}{Theorem}[section]

\newcommand{\bx}{\mathbf{x}}

\newcommand{\bmu}{\boldsymbol{\mu}}

\makeatother

\begin{document}
\baselineskip .3in

\title{
An approximate Bayesian inference on propensity score estimation under  unit nonresponse }

\author{Hejian Sang \quad Jae Kwang Kim \textsuperscript{}{\footnote{\textsuperscript{}
		Department of Statistics, Iowa State University, Ames, IA, 50010, U.S.A}}}
	\maketitle
		
\begin{abstract}
	Nonresponse weighting adjustment using the response propensity score is a popular tool for handling unit nonresponse. 
	Statistical  inference after the nonresponse weighting adjustment is complicated because the effect of estimating the 
	propensity model parameter needs to be incorporated. 
	In this paper, we propose an approximate Bayesian approach to handle unit nonresponse with  parametric model assumptions on the response probability, but without model  assumptions for the outcome variable. The proposed Bayesian method is calibrated to the frequentist inference in that the credible region obtained from the posterior distribution asymptotically matches to the frequentist confidence interval obtained from the Taylor linearization method. 
	Unlike the frequentist approach, however, the proposed method  does not involve Taylor linearization. The proposed method can be extended to 
	handle over-identified cases in which there are more estimating equations than the  parameters. Besides, the proposed method can also be modified to handle nonignorable nonresponse.  Results from {two} simulation studies confirm the validity of the proposed methods, which are then applied to data from a Korean longitudinal survey.  
\end{abstract}


\newpage

\section{Introduction}\label{sec:intro}

Missing data is frequently encountered in many areas of statistics. 
When the response mechanism is missing at random in the sense of \cite{rubin1976inference}, 
one of the popular methods of handling missing data is to build a model for the response probability and use the inverse of the estimated response probability to construct weights for estimating parameters. Such weighting method is often called propensity score weighting and the resulting estimator is called propensity score estimator \citep{rosenbaum1983central}. The propensity score method has been well established in the literature. For examples, see  \cite{rosenbaum1987model}, \cite{flanders1991analytic},
\cite{robins1994estimation}, \cite{robins1995analysis},  
and \cite{kim2007nonresponse}. However, all the above researches were developed via the frequentist approaches.  Variance estimates using a Taylor linearization method or bootstrap are used for making frequentist inference.

In this paper, we are interested in developing Bayesian inference for propensity score estimation. One of the main advantages of Bayesian inference is that all the uncertainty in the estimation process can be built into the Bayesian computation automatically. That is, there is no need to conduct variance estimation separately in the Bayesian inference.  While the Bayesian method is widely used in many areas of statistics, the literature on  the Bayesian approach of propensity score estimation is sparse. \cite{an2010bayesian} proposed a Bayesian propensity score estimator jointly modeling the response mechanism and the outcome variable. However, specifying a correct outcome model is difficult under missing data and incorrect specification may lead to biased inference. 
\cite{mccandless2009bayesian} and \cite{kaplan2012two} also assumed joint models and obtained Bayesian credible regions in the context of casual inference. 

In this paper, our interest is in developing a new Bayesian approach without making any model assumptions on the outcome variable.  Since no parametric model assumptions on the outcome variable are used, there is no explicit likelihood function corresponding to $\theta$, the main parameter of interest. This makes it difficult to develop a Bayesian method for propensity score estimation. The challenge thus lies in properly incorporating the uncertainty in the propensity score estimation process into the Bayesian framework.

In this paper, we propose a novel approach featuring approximate Bayesian computation based on the summary statistics  \citep{beaumont2002approximate}.  The sampling distribution of summary statistics, which is the estimating equation itself, can be used to replace the likelihood part in deriving the posterior distribution. 
In the proposed Bayesian method,  the credible region obtained from the posterior distribution with a  flat prior asymptotically matches the frequentist confidence interval obtained from the Taylor linearization method. 
The computation for the proposed method is relatively simple and easy to understand.

To guarantee the consistency of estimators, the propensity score method requires the correct specification of the response model. To protect against model misspecification, \cite{robins1994estimation}, \cite{scharfstein1999adjusting}, and \cite{bang2005doubly} proposed the so-called doubly robust estimation, which requires either the propensity score model or the outcome regression model be correctly specified.  To achieve efficiency and robustness, we can add into the proposed Bayesian method additional estimating equations obtained from the auxiliary variables observed throughout the full sample. When we incorporate more equations than the parameters, the proposed  Bayesian method is modified to  solve the over-identifying situation. 



The rest of the paper is organized as follows. In Section \ref{sec:setup}, we introduce the basic setup of the general propensity score estimation problem. The proposed method is presented in Section \ref{sec:proposed}. The main result and asymptotic theory are discussed in Section \ref{sec: Asymp}. In Section \ref{sec:extension}, we developed a related method by extending our proposed method to incorporate the auxiliary information observed throughout the sample. We also presented how to incorporate data augmentation algorithm to handle nonignorable nonresponse in Section \ref{sec:NMAR}. The finite sample performance of the proposed methods is examined in an extensive simulation study in Section \ref{sec:simulation}. An application of the proposed methods to a longitudinal survey is presented in Section \ref{sec:application}. Some concluding remarks are made in Section \ref{sec:Discussion}. 

\section{Basic Setup}\label{sec:setup}
 Suppose that we are interested in estimating $\theta$ defined through $E\left\lbrace U\left(\theta;\bm X, Y \right) \right\rbrace=0 $ for some estimating function  $U(\theta ;  \bm X, Y)$.  Let $\left( \bm x_i, y_i\right) , i=1, \cdots, n, $ be independently and identically distributed (IID) realizations of random variable $\left(\bm X, Y \right) $.  Under complete data, we can obtain a consistent estimator of $\theta$ by  solving 
 \begin{eqnarray}
 \frac{1}{n}\sum_{i=1}^{n} U\left(\theta;\bm x_i, y_i \right)=0\label{eq:com_est}
 \end{eqnarray}
for $\theta$.  We assume that the solution to (\ref{eq:com_est}) is unique almost everywhere.

 Now, suppose that $\bm X$ is always observed and $Y$ is subject to missingness.  In this case, we can define the response indicator function for unit $i$ as 
 \begin{eqnarray}
\delta_i=\left\lbrace 
\begin{array}{ll}
1 & \text{if $y_i$ is observed}\\
0 & \text{otherwise.}
\end{array}\right. \nonumber\label{eq:def_delta}
\end{eqnarray}
We assume that $\delta_i$ are independently generated from a Bernoulli distribution with 
\begin{equation}
Pr( \delta_i =1\mid \bm x_i, y_i) = \pi\left(\phi; \bm x_i, y_i\right)  \label{eq:def_pi}
\end{equation}
for some parameter vector $\phi$ and $\pi(\cdot)$ is a known function.  In the logistic regression model, $\pi (x) = 1/\{1+ \exp (-x) \}$.

When missing data exist, we cannot apply  (\ref{eq:com_est}) directly.  Instead, using the parametric model for the response probability  in (\ref{eq:def_pi}), we can obtain the propensity score (PS) estimator of $\theta$ by the following two steps: 
\begin{description}
\item{[Step 1]} Compute the maximum likelihood (ML) estimator $\hat{\phi}$ of $\phi$. 
\item{[Step 2]} Compute the PS estimator of $\theta$ by solving 
 \begin{eqnarray}
 \frac{1}{n} \sum_{i=1}^{n} \frac{\delta_i}{\pi (\hat{\phi} ;\bm x_i, y_i) } U\left(\theta; \bm x_i, y_i \right)=0 \nonumber\label{eq:adjust eq}
 \end{eqnarray}
for $\theta$. 
\end{description}

The computation for the ML estimator of $\phi$ can be greatly  simplified if the response mechanism is Missing At Random (MAR) in the sense that 
 \begin{eqnarray}
 Pr\left(\delta=1|\bm x, y \right)=Pr\left( \delta=1|\bm x\right). \nonumber \label{eq:def_MAR}
 \end{eqnarray}
 In this case, the maximum likelihood estimator of $\phi$ can be obtained by finding the maximizer of 
 \begin{equation}
   L (\phi) = \prod_{i=1}^n \left\{ \pi( \phi;\bx_i) \right\}^{\delta_i} \left\{ 1- \pi( \phi;\bx_i) \right\}^{1- \delta_i}  . 
   \label{eq:likeihood_phi}
   \end{equation}
     If MAR does not hold, parameter estimation is more complicated. Assuming a parametric model for $f_1 (y\mid \mathbf{x}) = f( y \mid \mathbf{x}, \delta=1)$, the ML estimator can be obtained by maximizing  $$ l_{obs} ( \phi) = \sum_{i=1}^n \delta_i \log \pi( \phi; \mathbf{x}_i, y_i  ) + \sum_{i=1}^n ( 1- \delta_i) \log \int \{ 1- \pi( \phi; \mathbf{x}_i, y_i  ) \} 
\hat{f}_1 ( y \mid \mathbf{x}_i) dy , $$
where $\hat{f}_1 ( y \mid \mathbf{x}_i)$ is an estimator for $f_1( y \mid \mathbf{x}_i)$. 
  \cite{riddles2016propensity} proposed an alternative computational tool that avoids computing the above integration using an EM algorithm.   


We shall first  present our proposed method under the MAR assumption. An extension to Not Missing At Random (NMAR) will be discussed in Section \ref{sec:NMAR}. 
Once the PS estimator $\hat{\theta}_{PS}$ of $\theta$ is obtained from the above two-step procedure, statistical inference for   
  $\theta$ can be made based on the asymptotic normality
 \begin{equation}
  \sqrt{n} ( \hat{\theta}_{PS} - \theta ) \stackrel{\mathcal{L}}{ \longrightarrow} N (0, \sigma^2) 
  \label{res2}
  \end{equation}
 for some   $ \sigma^2>0$, where $\stackrel{\mathcal{L}}{ \longrightarrow}$ denotes convergence in distribution.  See Chapter 5 of \cite{kim2013statistical} for a justification for (\ref{res2}).

Under the above setup, we shall introduce the proposed Bayesian approach to estimate the parameter and make inference from the posterior distribution. An advantage of the Bayesian approach is that we can incorporate the uncertainty in estimating $\phi$ into the Bayesian computation automatically.


\section{Proposed Method}\label{sec:proposed}

We now present the proposed Bayesian method in the case of MAR. Under the parametric model assumption (\ref{eq:def_pi}), the  likelihood function for $\phi$ is given in  (\ref{eq:likeihood_phi}).   From the likelihood function, we can derive the score function for $\phi$ as
  \begin{eqnarray}
  U_1\left( \phi\right)= \frac{1}{n}\sum_{i=1}^{n}\left\lbrace \frac{\delta_i}{\pi\left( \phi;\bm x_i\right)}-\frac{1-\delta_i}{1-\pi\left( \phi;\bm x_i\right)}\right\rbrace \frac{\partial \pi\left( \phi;\bm x_i\right)}{\partial \phi}=:\frac{1}{n}\sum_{i=1}^{n}s\left( \phi;\bm x_i,\delta_i\right).  \label{eq:U1}
  \end{eqnarray}
If we define 
\begin{eqnarray}
U_2\left(\phi,\theta \right)=\frac{1}{n}\sum_{i=1}^{n} \frac{\delta_i}{\pi\left( \phi;\bm x_i\right)} U\left(\theta; \bm x_i, y_i \right) , \label{eq:U2}
\end{eqnarray}
the PS estimator $\hat{\theta}_{PS}$ of $\theta$ can be viewed as the solution to the joint estimating equations:  $U_1( \phi) = 0 $ and $U_2 ( \phi, \theta) = 0$. Taylor linearization can be used to obtain a consistent variance estimator of $\hat{\theta}_{PS}$. See Chapter 5 of Kim and Shao (2013) for more details.

To introduce the proposed Bayesian inference corresponding to $\hat{\theta}_{PS}$, we first define $\boldsymbol\zeta = (\theta, \phi)$ and 
\begin{eqnarray}
U_n( \boldsymbol\zeta) =\left(\begin{matrix}
U_1\left( \phi\right)\\
U_2\left(\phi,\theta \right) 
\end{matrix} \right)\nonumber .
\end{eqnarray}
 Instead of generating the posterior distribution from $p(\boldsymbol\zeta \mid \mbox{sample})$ directly , we use the posterior distribution $p( \boldsymbol\zeta \mid \hat{\boldsymbol{\zeta}})$ to approximate the posterior distribution $p(\boldsymbol\zeta \mid \mbox{sample})$, where $\hat{\boldsymbol{\zeta}}$ solves $U_n( \boldsymbol\zeta)=0$. Thus, we can consider 
\begin{equation}
 p( \boldsymbol\zeta \mid \hat{\boldsymbol{\zeta}} ) = \frac{ g( \hat{\boldsymbol{\zeta}} \mid \boldsymbol\zeta) \pi (\boldsymbol\zeta) }{ \int g( \hat{\boldsymbol{\zeta}} \mid \boldsymbol\zeta) \pi (\boldsymbol\zeta) d \boldsymbol\zeta } 
 \label{12a}
\end{equation}
as an approximate posterior distribution for $\boldsymbol\zeta$, where  $g( \hat{\boldsymbol{\zeta}} \mid \boldsymbol\zeta)$ is the sampling distribution of $\hat{\boldsymbol{\zeta}}$ and $\pi(\boldsymbol \zeta)$ is the prior distribution for $\boldsymbol\zeta$. However, finding the sampling distribution $g( \hat{\boldsymbol{\zeta}} \mid \boldsymbol\zeta)$ will involve Taylor linearization. 

To consider an alternative computation, instead of generating from $p( \boldsymbol\zeta \mid \hat{\boldsymbol{\zeta}} )$ in (\ref{12a}), we use a posterior distribution from 
\begin{equation}
p( \boldsymbol\zeta \mid U_n  )= \frac{ g\{ U_n({\boldsymbol\zeta})  \mid  \boldsymbol\zeta\} \pi (\boldsymbol\zeta) }{ \int 
g\{  U_n ({\boldsymbol\zeta})  \mid \boldsymbol\zeta \}  \pi (\boldsymbol\zeta) d\boldsymbol \zeta } ,
\label{13a}
\end{equation}
where $g\{  U_n({\boldsymbol\zeta})  \mid  \boldsymbol\zeta\} $ is the sampling distribution of $U_n (\boldsymbol\zeta)$. 
To generate samples from (\ref{13a}), 
we first make a transformation of the parameters, defined as $\boldsymbol\eta= E( U_n \mid \boldsymbol\zeta)$.   Thus, $ T: \boldsymbol\zeta \rightarrow \boldsymbol\eta$ is an one-to-one transformation of the parameter. We will generate $\bm \eta^*$ from $p(\bm \eta\mid U_n)$ first and then use $\bm \zeta^*=T^{-1}(\bm \eta^*)$ to obtain the posterior distribution values from (\ref{13a}).

Now, to compute $p(\bm \eta\mid U_n)$, first note that, under some regularity conditions, 
\begin{eqnarray}
\left[\sqrt n U_n|\bm\zeta\right]=\left[\sqrt n U_n|\bm\eta\right] \xrightarrow{\mathcal L}N\left(\bm \eta, \Sigma \right) , 
\label{asym2}
\end{eqnarray}
where  notation $\left[ \cdot\right] $ is used to denote the sampling distribution and $\stackrel{\mathcal{L}}{ \longrightarrow}$ denotes the convergence in distribution. Writing $\pi( \bm\eta) $ as a prior distribution of $\bm \eta$,    the posterior distribution of $\bm\eta$ given $ U_n$ can be expressed as 
\begin{eqnarray}
\left[ \bm\eta|U_n\right] \propto \left[ U_n|\bm\eta\right]  \pi\left( \bm\eta\right).\nonumber
\end{eqnarray}
If there is no information for the prior, we can use a flat prior for $ \bm\eta $. The sampling distribution $\left[ U_n|\bm\eta\right]$ serves 
the role of the likelihood function in the Bayesian inference. Using  (\ref{asym2}) 
and  a flat prior for $\bm \eta$, we obtain 
\begin{eqnarray}
\left[ \bm \eta \mid  U_n \right] \sim N\left( \mathbf 0, \Sigma/n\right) 
\label{14} 
\end{eqnarray}
as the posterior distribution, where a consistent estimator of $\Sigma$ is
\begin{eqnarray}
\hat\Sigma=\left(\begin{matrix}
n^{-1} \sum_{i=1}^{n}s( \hat\phi;\bm x_i)^{\otimes 2}  & n^{-1} \sum_{i=1}^{n} \delta_i \hat{\pi}_i^{-1} s ( \hat\phi;\bm x_i )U'(\hat{\theta};\bm x_i,y_i )  \\
 \mbox{symm.}  & n^{-1} \sum_{i=1}^{n} \delta_i \hat{\pi}_i^{-2} U(\hat{\theta};\bm x_i,y_i ) ^{\otimes 2} 
\end{matrix} \right) ,\nonumber \label{eq:hat_sigma}
\end{eqnarray}
where $\hat{\pi}_i = \pi ( \hat{\phi}; \mathbf{x}_i)$, $\hat \phi $ and $\hat{\theta}$ solve $U_n\left(\phi,\theta \right)=\bm 0 $, $\bm A^{\otimes 2} =\bm A \bm A'$ and $\bm A'$ represents the transpose of $\bm A$.
The details of the derivation are presented  in Appendix A.  After we obtain the posterior distribution of $\bm \eta$, we can use the  inverse transformation of $T$  to obtain  the posterior distribution of the original parameters.
The following algorithm describes how to generate  parameters from the posterior distribution of $\bm \zeta=(\phi, \theta) $:

\begin{description}
	\item{[Step 1]} Generate $\bm\eta^* $ from the posterior distribution 
	\begin{equation}
p( \bm\eta\mid U_n=0) \stackrel{\mathcal{L}}{\longrightarrow} 	N(\bm 0, \hat \Sigma/n ) ,
\label{18}
\end{equation}
	 where $\hat{\Sigma}$ is a consistent estimator of $Var( \sqrt n U_n) = \Sigma$ in (\ref{asym2}). 
	\item{[Step 2]} Solve $U_n\left(\bm\zeta \right)=\bm\eta^* $ for $\bm\zeta$ to  obtain $\bm\zeta^*$. 
\end{description}	
Steps 1--2 can be repeated independently to generate independent samples from the posterior distribution. The samples can be used to obtain the posterior distribution of the induced parameters.

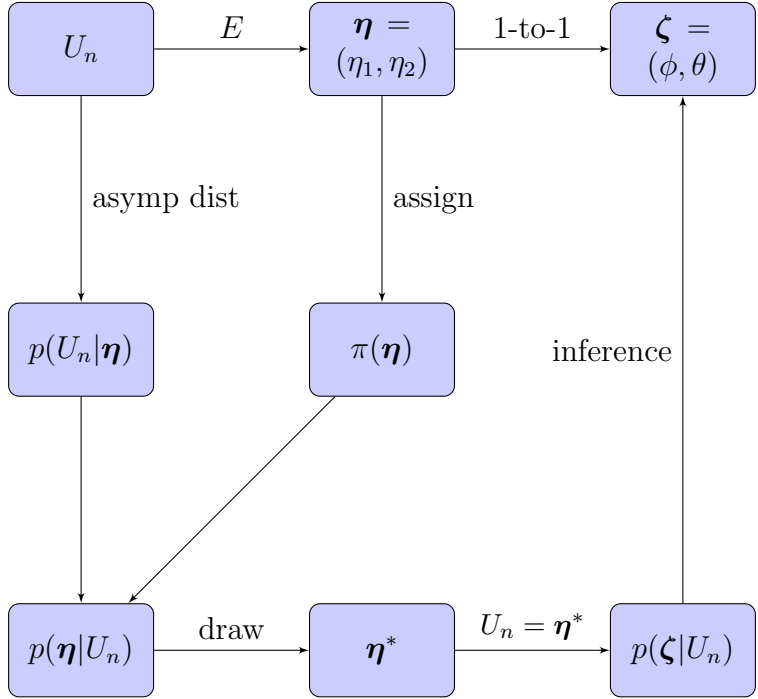
\begin{figure}[ht]
	\begin{center}
			\tikzstyle{decision} = [diamond, draw, fill=blue!20, 
			text width=5em, text badly centered, node distance=6cm, inner sep=0pt]
			\tikzstyle{block} = [rectangle, draw, fill=blue!20, 
			text width=4em, text centered, rounded corners, minimum height=3em]
			\tikzstyle{line} = [draw, -latex']
			\tikzstyle{cloud} = [draw, ellipse,fill=red!20, node distance=3cm,
			minimum height=2em]
			\begin{tikzpicture}[node distance = 4cm, auto]
			\node [block] (eq) {$U_n$};
			\node [block, right of=eq] (eta) {$\bm \eta=(\eta_1, \eta_2)$};
			\node[block, right of=eta] (zeta) {$\bm \zeta=(\phi,\theta)$};
			\node [block, below of=eta] (prior) {$\pi(\bm\eta)$};
			\node [block, below of=eq] (data) {$p( U_n|\bm\eta) $};
			\node [block, below of=data] (pos1) {$p(\bm{\eta}|U_n) $};
			\node[block, right of=pos1] (star) {$\bm\eta^*$};
			\node [block, right of=star] (pos2) { $p(\bm \zeta|U_n) $};
			\path [line] (eq) -- node {$E$}(eta);
			\path [line] (eta) -- node {1-to-1}  (zeta);
			\path [line] (eq) -- node {asymp dist} (data);
			\path [line] (eta) -- node {assign} (prior);
			\path [line] (data)-- (pos1);
			\path [line] (prior) --  (pos1);
			\path [line] (pos1) -- node {draw} (star);
			\path [line] (star)-- node {\small $U_n=\bm\eta^*$} (pos2);
			\path[line] (pos2)-- node {inference} (zeta);
			\end{tikzpicture}
			\caption{Proposed Bayesian propensity score method}\label{fig:1}
	\end{center}
\end{figure}	
 As we have illustrated before, the basic idea is that we use the posterior distribution of $p( \bm \zeta|\hat{ \bm{\zeta}}) $ to approximate the posterior distribution of $p\left(\bm \zeta|\text{sample} \right) $. This idea is 
similar in spirit to the Approximate Bayesian Computation of  \cite{soubeyrand2015weak}. Note that we do not use Taylor linearization to obtain the posterior distribution of $\theta$.  Instead, we use a transformation technique and generate the posterior distribution of  $p( \bm\eta |U_n)$ first. After we obtain the posterior distribution of $\bm\eta$, we use  the inverse transformation $T^{-1}: \bm\eta  \rightarrow \bm\zeta$  to obtain the posterior distribution of the original parameters.  The back-transformation plays the role of Taylor linearization in the frequentist approach. 
See Figure \ref{fig:1} for the illustration of the basic idea.
Some asymptotic properties are established in the next section.

\section{Asymptotic  Properties}\label{sec: Asymp}

 To establish the consistency of the parameter estimate and the interval estimate, we assume the following regularity conditions:  
\begin{description}
	\item{[C1]} As $n \rightarrow \infty$,  $U_n\left(\bm\zeta\right)\xrightarrow{}\bm \eta\left( \bm\zeta\right)$ in probability uniformly. That is $\sup_{\bm\zeta\in \boldsymbol{Z}}\|U_n\left(\bm\zeta\right)-\bm \eta\left( \bm\zeta\right)\|\xrightarrow{P}0$, where $\boldsymbol{Z}$ is the parameter space.\label{C1}
	\item{[C2]} The mapping $\bm\zeta\mapsto U_n\left(\bm\zeta \right)$ is continuous and has exactly one zero $\hat{\bm\zeta}$ with probability one as $n \rightarrow \infty$. 
	\label{C2}
	\item{[C3]} Equation  $\bm \eta\left( \bm\zeta\right)=0 $ has exactly one root at $\bm\zeta= \bm\zeta_0$. \label{C3}
	\item{[C4]} There exits a neighbor of $\bm\zeta_0$, denoted by $N_n\left(\bm\zeta_0 \right) $, on which with probability one all $U_n\left( \bm\zeta\right) $ are continuously differentiable and the Jacobian $\partial U_n\left(\bm \zeta\right) /\partial \bm\zeta$ converge uniformly to a non-stochastic limit which is non-singular. Here, $N_n\left(\bm\zeta_0 \right) $ is a ball with center $\bm\zeta_0$ and radius $r_n$, where $r_n$ satisfies $r_n\xrightarrow{}0$ and $r_n\sqrt n\xrightarrow{}\infty$. Also, we assume that $\partial^2 U_{n,j}\left(\bm\zeta \right) / \partial \bm\zeta \partial \bm\zeta'$ is finite for each entry for $j=1,2,\cdots,p$ and  with probability one as $n\xrightarrow{}\infty$.\label{C4}
	\item{[C5]} For any $\bm\zeta\in N_n\left(\bm\zeta_0 \right) $, 
	\begin{eqnarray}
	\sqrt n\left(  U_n\left(\bm\zeta \right)-\bm \eta\left( \bm\zeta\right) \right) \xrightarrow{\mathcal{L}} N\left(0, \Sigma\left( \bm\zeta\right)  \right)  
	\label{12}
	\end{eqnarray}
	holds for some $\Sigma ( \bm\zeta)=Var\left\lbrace  \sqrt n U_n\left(\bm\zeta \right)|\bm\zeta\right\rbrace >0$ that is independent of $n$.  \label{C5}	
\end{description}
As long as the samples satisfy some moment conditions, condition [C1] holds. Condition [C2] and [C3] are used to make sure that the solutions of estimating equation $U_n$ and estimating function $\boldsymbol\eta$ exist and are unique to avoid the model non-identifiability problem. The condition [C4] regulates the derivatives of the estimating equation to make sure that the variance converges. Condition [C5] provides the asymptotic distribution for the estimating equation. Under the above conditions,  we can obtain
	\begin{eqnarray}
	\sqrt n \left(\hat{\bm\zeta}-\bm\zeta_0 \right)\xrightarrow{\mathcal{L}} N\left(0,A( \bm\zeta_0) ^{-1} \Sigma\left(\bm\zeta_0 \right) A'(\bm\zeta_0)^{-1} \right) 
	\label{13}
	\end{eqnarray}
	where $A(\bm\zeta) = \partial \bm \eta\left(\bm \zeta\right) / \partial \bm\zeta $. 

 We now make additional assumptions to establish the posterior consistency and convergence in distribution:

\begin{description}
	\item{ [C6]}  The  prior distribution $\bm \eta\mapsto\pi\left(\bm \eta\right) $  is positive and Lipschitz continuous over the parameter space.\label{C6}
	
	\item{[C7]} For any $\bm\zeta \in N_n\left(\bm\zeta_0 \right) $,  the variance estimator   $\hat{\Sigma}\left(\bm\zeta \right) $ in (\ref{18})   satisfies $  \hat{\Sigma}  \left(\bm\zeta \right)  = 
	\Sigma \left(\bm\zeta \right) \left\lbrace 1+o_p(1) \right\rbrace  $.\label{C7}

	\item{[C8]}  For any $\bm\zeta \in N_n\left(\bm\zeta_0 \right) $, the mapping   $\bm\zeta\mapsto\left| \Sigma \left(\bm\zeta \right) \right|^{-1}$ is Lipschitz continuous. Also, the mapping  $\bm\zeta\mapsto x'\left\lbrace  \Sigma\left(\bm\zeta \right) \right\rbrace^{-1} x$ is Lipschitz continuous 
	in the sense that there exists a constant $C\left(x \right) $ satisfying $\left\|x'\left\lbrace  \Sigma\left(\bm\zeta_1 \right) \right\rbrace^{-1} x-x'\left\lbrace  \Sigma\left(\bm\zeta_2 \right) \right\rbrace^{-1} x \right\|\leq C\left( x\right) \left\|\bm\zeta_1-\bm\zeta_2 \right\|  $, for any $\bm\zeta_1,\bm\zeta_2\in N_n\left(\bm\zeta_0 \right) 	$, for all $x \in   \mathbb R^p$, where $p=\dim\left(\boldsymbol{Z} \right) $.  And $C\left(x \right) $ is also Lipschitz continuous.
	\item{[C9]} $\bm\zeta\mapsto U_n\left(\bm\zeta\right) $ and $\bm\zeta\mapsto \bm \eta\left( \bm\zeta\right) $ are one to one functions for any $\bm\zeta\in N_n\left(\bm\zeta_0 \right) $. Also  $\bm\zeta\mapsto \bm \eta\left( \bm\zeta\right) $  is Lipschitz continuous.\label{C9} 
\end{description}
 Condition [C6] is a common assumption for the prior and the flat prior satisfies this condition.   Condition [C7] requires the variance estimator  to be consistent. Conditions [C8] to [C9] are the sufficient conditions for the posterior distribution to be approximated by the proposed method. All the conditions can be easily satisfied if we assume variance estimate is continuous and has bounded eigenvalues. 

\begin{theorem}\label{theorem3}
Let $\hat{\bm\zeta}$ be the solution to $U_n( \bm\zeta)=0$. 
	Under (C1)--(C9), the posterior distribution  $p( \bm\zeta \mid U_n=0)=p(\bm \zeta|\hat{\bm \zeta})$, generated by the two-step method in Section \ref{sec:proposed}, satisfies
	\begin{eqnarray}
	p( \bm\zeta|\hat{\bm\zeta}) \xrightarrow{} \phi_{\hat{\bm\zeta},Var(\hat{\bm\zeta})}\left( \bm\zeta\right) \label{21} \\
	p \lim_{n\xrightarrow{}\infty} \int_{N_n\left(\bm\zeta_0 \right)} \phi_{\hat{\bm\zeta},Var(\hat{\bm\zeta})}\left( \bm\zeta\right) d\bm\zeta =1 , \label{22}
	\end{eqnarray}
	where $\phi_{\hat{\bm\zeta},Var(\hat{\bm\zeta} )}\left( \cdot\right) $ is the density of the normal distribution with mean $\hat{\bm\zeta}$ and variance $Var(\hat{\bm\zeta} )$.
\end{theorem}
The proof is shown in Appendix B. Result (\ref{21}) is a convergence of the posterior distribution to normality and result (\ref{22}) is the posterior consistency. By (\ref{21}), the confidence region using the proposed Bayesian method is asymptotically equivalent to the frequentist confidence region based on asymptotic normality of $\hat{\bm\zeta}$. Thus, our proposed Bayesian method is calibrated to frequentist inference.


To construct a level $\alpha$ confidence region,  let $k^* (\alpha)$ be the largest value of $k$ such that 
$$ Pr \{   \bm\zeta: p( \bm\zeta \mid \hat{\bm\zeta}) \ge k  \} = 1 - \alpha . $$
The level-$\alpha$ Bayesian High Posterior Density (HPD) confidence region \citep{chen1999monte} using $k^*$ is
		\begin{eqnarray}
		C^*(\alpha)=\left\lbrace \bm\zeta:
		p( \bm\zeta \mid \hat{\bm\zeta}) \ge k^* (\alpha) 		
		 \right\rbrace .
		 \label{23} \nonumber
		\end{eqnarray}
We can show that $\int_{ \hat C^*\left(\alpha \right)} p( \bm\zeta \mid \hat{\bm\zeta}) d\bm\zeta \xrightarrow{}1-\alpha$ in probability, where $\hat C^*(\alpha)$ is the confidence region from Monte Carlo samples, which are generated from the approximate target posterior distribution. See \cite{hyndman1996computing}. \label{claim_corollary} 


\section{Optimal Estimation 
}\label{sec:extension}

We now extend the proposed method to incorporate additional information from the full sample. Note that the PS estimator applied to $\bmu_x = E( X)$ can be computed as the solution to 
$$  \sum_{i=1}^n \frac{ \delta_i}{ \pi( \hat{\phi}; \bm{x}_i) } (\bm{x}_i - \bmu_x) =0$$
which is not necessarily equal to $\hat{\mu}_{x,n}=n^{-1} \sum_{i=1}^n \bm x_i$. Including this extra information in the propensity score estimation, if done properly,  will improve the efficiency of the resulting PS estimator. In the frequentist propensity score method, incorporating  such extra information can be implemented 
by Generalized Method of Moments and it is sometimes called optimal PS estimation. See  \cite{cao2009improving}, \cite{zhou2012efficient} and 
\cite{imai2014covariate}.

To include such extra  information, we may add 
\begin{eqnarray}
U_3\left(\phi,\bm \mu_x \right) &=& \frac{1}{n} \sum_{i=1}^{n} \frac{\delta_i}{\pi\left(\phi;\bm x_i \right) }\left( \bm x_i-\bm \mu_x\right)\nonumber\label{eq:U_3}\\
U_4\left(\bm \mu_x \right)&=&\frac{1}{n} \sum_{i=1}^{n}\left(\bm x_i-\bm \mu_x \right) \nonumber\label{eq:U_4}
\end{eqnarray}
in addition to the original estimating equations based on $U_1(\phi)$ and $U_2( \phi, \theta) $ in (\ref{eq:U1}) and (\ref{eq:U2}), respectively. Note that we cannot directly apply the proposed two-step method in Section \ref{sec:proposed} in this case  because
there are more estimating equations than the parameters and the transformation 
$$ (\bmu_x, \phi, \theta) \rightarrow (\bm\eta_1, \eta_2, \bm\eta_3, \bm\eta_4)$$ 
is not one-to-one, where
\begin{eqnarray}
\left( \begin{matrix}
\bm \eta_1 \\
\bm \eta_2 \\
\bm \eta_3\\
 \bm \eta_4
\end{matrix}\right) =E\left\lbrace \left. \left( \begin{matrix}
U_1 (\phi) \\
U_2 ( \theta, \phi) \\
U_3\left(\phi,\bm \mu_x \right)\\
U_4\left(\bm \mu_x \right)
\end{matrix}\right)\right|\bmu_x, \phi, \theta \right\rbrace \nonumber.
\end{eqnarray}


To solve this problem, instead of using the two-step method involving generation of $\bm \eta^*$ first from (\ref{18}), we consider a direct sampling method that generates $\bm \psi^*=( \bmu_x^*, \phi^*, \theta^*)$ from the posterior distribution of $\bm \psi= ( \bmu_x, \phi, \theta)$ given the observed data directly.  To formally describe the procedure, first define 
$$U_n (\bm \psi) =\left(U_1'(\phi),U_2(\phi,\theta),U_3'(\phi,\bmu_x), U_4'(\bmu_x) \right)' . $$ 
 Under some regularity conditions, we can obtain
 \begin{equation}
[U_n|\boldsymbol \psi]\sim N(\mathbf 0, \Sigma(\boldsymbol\psi)/n)
\label{26}
\end{equation}
for sufficiently large $n$, where $\Sigma( \boldsymbol \psi )= Var \left\{ \sqrt{n} U_n ( \bm \psi)   \mid \bm \psi  \right\} $.  
Using (\ref{26}) as the sampling distribution $g(U_n|\boldsymbol \psi)$ of $U_n$ and  using a prior    $\pi(\boldsymbol \psi)$ for  $\boldsymbol \psi$, the posterior distribution of $\boldsymbol \psi$  can be written as 
  \begin{equation}
p\left(\boldsymbol\psi|U_n \right)=
\frac{g(U_n|\boldsymbol \psi)\pi(\boldsymbol \psi)}{\int g(U_n|\boldsymbol \psi)\pi(\boldsymbol \psi) d\boldsymbol \psi} . 
\label{30} 
\end{equation} 
Note that we can still use the approximate normality of $U_n$ to play the role of the likelihood function in the approximate Bayesian analysis. Note that even if the prior distribution is normal, the posterior distribution  in (\ref{30}) 
is no longer normal.

To obtain the posterior draws from (\ref{30}), 
we can use a Monte Carlo method based on a version of  Metropolis-Hastings algorithm (e.g. \citet{chib1995understanding}). The computation details of the Monte Carlo method for generating samples from  (\ref{30}) are presented in Appendix C. 

Note that, in generating samples from (\ref{30}),  
 the number of estimating equations is allowed to be greater than the number of parameters. Therefore, the proposed method is quite flexible in the sense that it can be applied to over-identified situations. Since the point estimator is asymptotically equivalent to the optimal PS estimator, the proposed method can thus be called optimal Bayesian PS (OBPS) method.

\section{Nonignorable nonresponse}\label{sec:NMAR}

We now consider an application of the proposed Bayesian method to nonignorable nonresponse. 
Under the setup of Section \ref{sec:setup}, we first assume a parametric model for the response mechanism \begin{eqnarray}
Pr(\delta_i=1|\bm x_i, y_i)=\pi(\phi;\bm x_{i1}, y_i),\label{eq:response_NMAR}
\end{eqnarray}
where  $\pi\left( \cdot\right) $ is known up to $\phi$ and $\bm x_i=(\bm x_{i1}, \bm x_{i2})$.  The auxiliary variable $\bm x_{i2}$ is often called the response instrumental variable to avoid the non-identifiable problem in \cite{wang2014instrumental}. In addition, we assume a parametric model for the respondents' outcome model 
\begin{equation}
 f( y_i \mid \mathbf{x}_i , \delta_i=1) = f_1 ( y_i \mid \mathbf{x}_i; \gamma) \label{d2}
 \end{equation}
for some $\gamma$. Using (\ref{eq:response_NMAR}) and (\ref{d2}), we can obtain
the following prediction model for the nonrespondents:  
\begin{eqnarray}
f\left(y|\bm x, \delta=0; \gamma, \phi \right)=f\left(y|\bm x, \delta=1; \gamma  \right)\frac{O\left(\bm x_1, y;\phi \right) }{E\left\lbrace O\left(\bm x_1, y;\phi \right)|\bm x, \delta=1 \right\rbrace },\label{eq:link}
\end{eqnarray}
where $O\left(\bm x_1, y;\phi \right)=Pr\left(\delta=0|\bm x_1, y \right) /Pr\left(\delta=1|\bm x_1, y \right)$, $ f\left(y|\bm x, \delta=1 \right)$. If $\pi ( \phi; \mathbf{x}_{i1}, y_i)$ follows a logistic regression model such as 
$ \pi(\phi;\bm x_{i1}, y_i) = \{1+ \exp ( x_{i1} \phi_1 + y_i \phi_2) \}^{-1}$  
then $O\left(\bm x_1, y;\phi \right)=\exp ( - \phi_2 y)$. See \cite{kim2011semiparametric} for more discussion of the prediction model (\ref{eq:link}). 

If  $y_i$ were available throughout the sample,  we could  use 
\begin{eqnarray*}
S_{1}\left( \gamma\right)&:=&\frac{1}{n} \sum_{i=1}^{n}\delta_i s_1\left(\gamma;\bm x_i, y_i \right) \\ 
S_2\left(\phi \right) &:=& \frac{1}{n}\sum_{i=1}^{n}s_2\left( \phi;\delta_i,\bm x_{1i}, y_i\right) \label{eq:com_score_phi},\\
 U(\theta)&=& \frac{1}{n}\sum_{i=1}^{n}U(\theta;\bm x_i, y_i),\label{eq:theta_eq}
\end{eqnarray*}
as the estimating functions for $\bm\zeta=( \gamma,\phi, \theta)$, where $s_1(\gamma)$ is the score function of $\gamma$ with $s_1( \gamma; x_i, y_i) =\partial \log f\left(y_i|\bm x_i,\delta_i=1;\gamma \right) / \partial \gamma$ and $S_2\left(\phi \right)$ is the score function of $\phi$. 
 Writing  the joint estimating equations as $U_n(\bm\zeta)=\left( S'_1(\gamma), S'_2(\phi), U(\theta)\right)' $ and $\boldsymbol{\eta}=E\left\lbrace U_n(\bm\zeta)\mid \bm\zeta \right\rbrace $, the following two-step method can be used to generate the posterior samples of $\bm\zeta$. 
\begin{description}
\item{[Step 1]} Generate $\boldsymbol{\eta}^*$ from the approximate posterior distribution using $p(\boldsymbol{\eta} \mid U_n (\bm\zeta)  )$. Under a flat prior for $\boldsymbol{\eta}$,   the posterior distribution of $\boldsymbol{\eta}$ can be obtained  
 as a multivariate normal distribution with mean $\bm 0$ and variance $\Sigma/n$. A consistent estimator of $\Sigma$ is 
\begin{eqnarray}
\hat{\Sigma}= \frac{1}{n}\sum_{i=1}^{n}\left(\begin{matrix}
\delta_i s_1\left(\hat \gamma;\bm x_i, y_i \right)\\
s_2( \hat \phi;\delta_i,\bm x_{1i}, y_i)\\
U(\hat \theta;\bm x_i, y_i)
\end{matrix} \right)^{\otimes 2},\nonumber
\end{eqnarray}
where $\hat{\bm\zeta}= (\hat \gamma, \hat \phi,\hat \theta)$ is the solution to  $U_n(\bm\zeta)=\bm 0$ under complete response. 
\item{[Step 2]} The posterior values of $\bm\zeta$ can be obtained by solving $U_n( \bm\zeta) =  \boldsymbol{\eta}^*$ for $\bm\zeta$. 
\end{description}

Now, to implement the proposed Bayesian method under missing data, we can use 
Data Augmentation (DA) method of \cite{tanner1987calculation}.  The DA algorithm consists of I-step and P-step. In  I-step, the imputed values of $y_i$ are generated from the prediction model using the current parameter values. In P-step, the posterior values of the parameters are generated from the above two-step method using the current imputed data. To formally describe the proposed method, define $X_n=\left\lbrace\bm x_1, \cdots, \bm x_n \right\rbrace $, $\bm \delta_n=\left\lbrace \delta_1, \cdots, \delta_n\right\rbrace $ and $Y_n=(Y_{\text{obs}}, Y_{\text{mis}})$, where  $Y_{\text{obs}}$ and $ Y_{\text{mis}}$ are the observed and missing part of $Y_n=(y_1, \cdots, y_n)$, respectively. 
 The proposed  DA algorithm  can be described as follows:  
\begin{description}
	\item [\textbf{I-step}:] Given current parameter values $\bm\zeta^*$, generate imputed values  $Y^*_{\text{mis}}$ from the prediction model  (\ref{eq:link}) evaluated at the current parameter values. 	
	\item[\textbf{P-step}:] Using the current imputed data, apply the above two-step method of generating the parameter values $\bm\zeta^*$ from $p( \bm\zeta \mid U_n^* (\bm\zeta))$, where $U_n^* (\bm\zeta)=U_n(\bm \zeta; Y_{\text{obs}}, Y^*_{\text{mis}})$.
	\end{description}
	The two steps are iteratively computed  until some convergence criterion is satisfied. Once the posterior values of $\bm\zeta^*$ are obtained,  the posterior values of $\theta^*$ can be used to perform  Bayesian inference for $\theta$. 
	To explain  the proposed method further, denote $p_U (\bm \zeta \mid X_n,Y_n,\bm \delta_n) = p( \bm\zeta \mid U_n)$ to emphasize that $U_n(\bm\zeta)$ is a function of $Y_n$.  The \textbf{I-step} of the proposed method is to generate $Y_{\text{mis}}$ from the posterior predictive distribution of $Y_{\text{mis}}$ by
\begin{eqnarray}
f(Y_{\text{mis}}|X_n, Y_{\text{obs}},\boldsymbol{\delta}_n)=\int f(Y_{\text{mis}}|X_n, \bm\zeta )p_U ( \bm\zeta \mid X_n,  Y_{obs},  \bm\delta_n   )  d\bm\zeta, \nonumber
\end{eqnarray}
where 
\begin{eqnarray}
p_U(\bm\zeta|X_n, Y_{\text{obs}},\boldsymbol{\delta}_n)=\int p_U (\bm\zeta |X_n, Y_n,\boldsymbol{\delta}_n)f(Y_{\text{mis}}|X_n, Y_{\text{obs}},\boldsymbol{\delta}_n)dY_{\text{mis}}\nonumber
\end{eqnarray}
is generated from  \textbf{P-step}.  After convergence, the DA algorithm generates  $\zeta$ from the posterior density 
$$ p_U(\bm\zeta|X_n, Y_{\text{obs}},\boldsymbol{\delta}_n)= \frac{ \int g( U_n \mid \bm \zeta ) \pi ( \bm  \zeta) d Y_{\text{mis}} }{
\int \int g( U_n \mid \bm \zeta ) \pi ( \bm \zeta) d Y_{\text{mis}} d \bm \zeta } . 
$$
 \section{Simulation Study}\label{sec:simulation}

 We perform two limited simulation studies to validate our theory and to check the robustness of our proposed methods. In the first simulation, the proposed method is evaluated under ignorable response mechanism. In the second simulation, the proposed method is applied to some nonignorable nonresponse mechasnism.

 \subsection{Simulation Study One}
  The first simulation study can be described as a $ 3 \times 4$ factorial design, where the factors are outcome regression model  for $E( y \mid \mathbf{x})$ and the response mechanism.

 For the outcome regression models, we use $y=m(x_1, x_2) + e$ with three different mean functions given by 
 \begin{eqnarray}
\begin{array}{ll}
\text{Function 1:} & m_1(\mathbf x)=2x_1+3x_2-20\\
\text{Function 2:} & m_2(\mathbf x)=0.5(x_1-2)^2+x_2-2\\
\text{Function 3:} & m_3(\mathbf x)=0.1\exp(0.1x_1-0.2)+3x_2+c_3 \end{array},\nonumber
\end{eqnarray} 
where $c_3$ is chosen to give the same values for $E(y)$ in different mean functions.  The explanatory variables 
$(x_1,x_2)^T$ are generated from $N(\bmu, \Sigma_{x}) $, with $\bmu_x=\left( 2,8\right)^T $ and $\Sigma=\mbox{diag} \{ 4, 8\}$. The error distribution is $e\sim N(0,\sqrt{|x_1|+1})$. 


For the response mechanism, we use four different response mechanisms. In the first response mechanism (R1), the response indicator function $\delta_i$ are independently generated from a Bernoulli distribution with probability 
\begin{eqnarray}
p_i\left( \phi_0,\phi_1\right)=\frac{\exp(\phi_0+\phi_1x_{i1})}{1+\exp(\phi_0+\phi_1x_{i1})} 
\end{eqnarray}
with  $(\phi_0,\phi_1)=(0.1,0.4)$, which  makes  the overall response rate  approximately equal to 70\%. In the second response mechanism (R2), we use the sample logistic regression model with  $(\phi_0,\phi_1)=(-1.2,0.15)$, which leads to about 30\% response rate. In the third response mechanism (R3), the  response indicator function $\delta_i$ are independently generated from a Bernoulli distribution with probability 
\begin{eqnarray}
p_i(\phi_0,\phi_1)=\Phi(\phi_0+\phi_1x_{i1})
\end{eqnarray}
where $\Phi(\cdot)$ is the cumulative distribution function of the standard normal distribution and   $(\phi_0,\phi_1)=(0,0.28)$, which leads to about 70\% response rate. In the fourth response mechanism (R4), we use the same probit model with  $(\phi_0,\phi_1)=(-0.7,0.1)$ to make the response rate near to 30\%.

For each of the $12=3\times 4$ simulation setup, we generate random samples of  size $n=500$ independently $B=2,000$ times. From  each realized sample, we  specify a logistic regression model   $$ Pr(\delta_i=1|\mathbf x_i,y_i)=\frac{\exp(\phi_0+\phi_1x_{i1}+\phi_2x_{i2})}{1+\exp(\phi_0+\phi_1x_{i1}+\phi_2x_{i2})}=:\pi(\phi;\mathbf x_i)$$ 
as the response model. Thus, in R3 and R4, the response model is incorrectly specified. 

For each Monte Carlo sample, we use the following four methods of inference for $\theta=E(y)$:

\begin{itemize}
	\item [1.] PS: Frequentist approach based on Taylor linearization.  The point estimator 
	 $(\hat \theta_{PS}, \hat \phi)$ 
	is computed from  
	  \begin{eqnarray}
	U_{PS}(\theta, \phi)=\frac{1}{n}\sum_{i=1}^{n}\frac{\delta_i}{\pi(\phi;\mathbf x_i)}(y_i-\theta)=0\nonumber\\
	S(\phi)=\frac{1}{n}\sum_{i=1}^{n}\{\delta_i-\pi(\phi;\mathbf x_i)\}(1,\mathbf x'_i)'=\mathbf 0 . \nonumber
	\end{eqnarray}
	The confidence intervals are constructed by $\hat\theta_{PS}\pm 1.96 \sqrt{\hat V_{PS}}$, where $\hat V_{PS}$ is obtained by the Taylor linearization method. 
			\item [2.] Bayesian PS (BPS): Apply the proposed Bayesian method based on  the joint estimating functions 
	\begin{eqnarray}
	U_1(\phi)=\frac{1}{n}\sum_{i=1}^{n}\left\lbrace \delta_i-\pi(\phi;\mathbf x_i)\right\rbrace (1,\mathbf x'_i)' \label{32}\\
	U_2(\phi, \theta)=\frac{1}{n}\sum_{i=1}^{n}\frac{\delta_i}{\pi(\phi;\mathbf x_i)}(y_i-\theta)\label{33} 
	\end{eqnarray}
	 	The estimators for $ \phi, \theta$ are obtained by the median of the draws from the approximate posterior distribution. The confidence interval can be constructed by HPD region introduced in Section \ref{sec: Asymp}.
	\item[3.] Optimal PS (OPS):  Generalized method of moments using 	\begin{eqnarray}
	U_3(\phi, \mu_x)=\frac{1}{ n}\sum_{i=1}^{n}\frac{\delta_i}{\pi(\phi;\mathbf x_i)}(\mathbf x_i-\bmu_x)\nonumber\\
	U_4(\mu_x)=\frac{1}{ n}\sum_{i=1}^{n}(\mathbf x_i-\bmu_x)\nonumber
	\end{eqnarray}
	in addition to (\ref{32}) and (\ref{33}).  If we denote $U_n(\mu_x, \phi, \theta)=(U'_1, U_2, U'_3, U'_4)'$, then the OPS estimator is obtained by minimizing $U_n^T W^{-1}U_n$, where $W=Var(U_n)$. See Section 5.4 of \cite{kim2013statistical}. 	
	
	\item [4.] OBPS:  Optimal Bayesian PS method discussed in Section 5 using the same estimating functions $U_1(\phi)$,  $U_2(\phi, \theta)$, $U_3(\phi, \mu_x)$, and $U_4(\mu_x)$. 		The point estimators for $\bmu_x, \phi, \theta$ are obtained by the median of the draws from the approximate posterior distribution.
	 The confidence intervals can be constructed by the HPD region, introduced in Section \ref{sec: Asymp}.
\end{itemize}

For each of the four methods, 95\% confidence intervals for $\theta$ are computed from Monte Carlo samples.

\begin{table}[ht]
	\centering
	\caption{Simulation results: ``m" denotes mean function,  ``c\_p" is the coverage probability for the corresponding confidence interval,  ``CI length" is the average length of the confidence intervals.}\label{tbl:model1}
	\begin{tabular}{l|l|lrr|l|l|lrr}
		\hline
		Response  & m & method & c\_p & CI & Response & m & method & c\_p &    CI\\ 
		mechanism & & &  &  length & mechanism & &  & &  length \\
		\hline
		\multirow{9}{*}{R1} & \multirow{3}{*}{$m_1$} & PS & 0.95 & 1.83 & \multirow{9}{*}{R3} & \multirow{3}{*}{$m_1$} & PS & 0.95 & 1.86 \\ 
		&  & BPS & 0.95 & 1.84  & & & BPS & 0.95 & 1.87 \\ 
		&  & OPS & 0.95 & 1.78  & & & OPS & 0.95 & 1.78 \\
		&  & OBPS & 0.95 & 1.78  & & & OBPS & 0.94 & 1.78 \\ 
		\cline{2-5}\cline{7-10}
		&  \multirow{3}{*}{$m_2$} & PS & 0.94 & 0.88 & & \multirow{3}{*}{$m_2$} & PS & 0.94 & 0.89\\ 
		&  & BPS & 0.94 & 0.88 & & & BPS & 0.94 & 0.89 \\ 
		&  & OPS & 0.94 & 0.79 & & & OPS & 0.93 & 0.79  \\
		&  & OBPS & 0.94 & 0.80 & & & OBPS & 0.94 & 0.80 \\
		\cline{2-5}\cline{7-10}
		&  \multirow{3}{*}{$m_3$} & PS & 0.95 & 1.56  & & \multirow{3}{*}{$m_3$} & PS & 0.94 & 1.58\\ 
		&  & BPS & 0.94 & 1.56  & & &  BPS & 0.94 & 1.58\\ 
		&  & OPS & 0.95 & 1.53  & & & OPS & 0.95 & 1.53 \\ 
		&  & OBPS & 0.94 & 1.52 & & & OBPS & 0.94 & 1.52 \\
		\hline
		\multirow{9}{*}{R2} & \multirow{3}{*}{$m_1$} & PS & 0.95 & 1.96 & \multirow{9}{*}{R4} & \multirow{3}{*}{$m_1$} & PS & 0.95 & 1.95  \\ 
		&  & BPS & 0.96 & 2.00 & & & BPS & 0.95 & 1.99 \\
		&  & OPS & 0.95 & 1.83 & & & OPS & 0.94 & 1.82\\  
		&  & OBPS & 0.95 & 1.83  & & & OBPS & 0.95 & 1.82 \\ 
		\cline{2-5}\cline{7-10}
		&  \multirow{3}{*}{$m_2$} & PS & 0.95 & 1.16  & & \multirow{3}{*}{$m_2$} & PS & 0.94 & 1.13 \\ 
		&  & BPS & 0.95 & 1.16 & & & BPS & 0.95 & 1.13 \\ 
		&  & OPS & 0.94 & 0.91 & & & OPS & 0.93 & 0.90 \\ 
		&  & OBPS & 0.95 & 0.97 & & & OBPS & 0.95 & 0.95 \\
		\cline{2-5}\cline{7-10}
		&  \multirow{3}{*}{$m_3$} & PS & 0.95 & 1.68  & & \multirow{3}{*}{$m_3$} & PS & 0.95 & 1.67\\ 
		&  & BPS & 0.95 & 1.72 & & & BPS & 0.95 & 1.70 \\ 
		&  & OPS & 0.95 & 1.59 & & & OPS & 0.95 & 1.58  \\ 
		&  & OBPS & 0.95 & 1.58 & & & OBPS & 0.95 & 1.57 \\
		\hline
	\end{tabular}
\end{table}


Table \ref{tbl:model1} presents the simulation results,  coverage probabilities and average  lengths of confidence intervals (CI),  for the four methods.
 Overall, all the coverage probabilities are approximately 95\%, which  validates our proposed methods BPS and OBPS. For R1 and R2, we have a correctly specified model for the response mechanism. For R1, which has high response rate 70\%, both BPS and OBPS methods  provide valid confidence intervals with correct coverage rates. Comparing the average length of confidence intervals, we can see that PS and BPS methods have approximately equal average CI lengths and OPS and OBPS have approximately equal average CI lengths, which confirms the asymptotic equivalence of the two methods. That is, our proposed Bayesian methods are calibrated to the frequentist inference.  The same conclusion can be obtained for R2, which has much lower response rates. For different regression mean functions, we find that both OPS and OBPS methods  achieve more efficiency gains when the regression model is not linear and the response rate is low. For the probit response mechanism (R3 and R4), 
 BPS and OBPS still  provide valid confidence intervals with correct coverages. Thus, the proposed method seems to be robust against model misspecification of the response model. 


\subsection{Simulation Study Two}

In the second simulation study, we consider an extension of the proposed method to nonignorable nonresponse. 
In the simulation, we generate the covariate variable $x\sim N(0, 0.5)$ and  use the outcome regression model  $y=m(x)+e$ to generate $y$, where $e\sim N(0,1)$. We consider three different mean functions $m(x)$, which are specified as $m_1(x)=-1+2x$, $m_2(x)=-1.25+2x+0.5x^2$ and $m_3(x)=-1+8\sin(x)$.

We use two different mechanisms to generate the response indicators. 
The response indicator function $\delta_i$ are independently generated from Bernoulli distribution with the  probability for $\delta_i=1$ equal to 
\begin{eqnarray}
p_i(\phi_0,\phi_1)=\left\lbrace 
\begin{array}{ll}
\{1+\exp(-\phi_{10}-\phi_{11}y_i)\}^{-1} &\mbox{for } \mathcal R_1 \\
\Phi(\phi_{20}+\phi_{21}y_i) & \mbox{for }  \mathcal R_2 , 
\end{array}\right. 
\end{eqnarray}
where $(\phi_{10},\phi_{11})=(0.8,-0.2), (\phi_{20}, \phi_{21})=(0.5,-0.1)$ and $\Phi(\cdot)$ is cumulative distribution function of the standard normal distribution. The overall response rates are approximately around 70\%. Thus, we have $3 \times 2$ setup for the simulation study. 

For each simulation setup, $n=500$ samples are generated independently for  2,000 times.  For each Monte Carlo sample, we apply the following methods to estimate $\theta=E(y)$:

\begin{itemize}
	\item [1.] Full sample method:  Use  $\hat \theta=\sum_{i=1}^{n}y_i/n$, which is computed as a benchmark for the comparison. 
	\item[2.] Complete-Case (CC) method: Estimate $\theta$ by removing nonresponse. That is, $\hat{\theta}_{CC}$ is obtained by solving $\sum_{i=1}^n \delta_i (y_i - \theta) =0$ for $\theta$. 
	\item[3.] \citet{kott2010using} (KC) method:  Assume the response model is 
	\begin{eqnarray}
	Pr(\delta_i=1\mid x_i, y_i)=\pi(\phi;y_i)=\frac{\exp(\phi_0+\phi_1y_i)}{1+\exp(\phi_0+\phi_1y_i)}.\label{eq:res_asumme}
	\end{eqnarray}
 The KC estimates are obtained by solving 
	\begin{eqnarray}
	\frac{1}{n}\sum_{i=1}^{n}\left\lbrace\frac{\delta_i}{\pi(\phi;y_i)}-1 \right\rbrace (1, x_i)'=\bm 0,\nonumber\\
	\frac{1}{n}\sum_{i=1}^{n}\frac{\delta_i}{\pi(\phi;y_i)}(y_i-\theta)=0.\nonumber
	\end{eqnarray}
	\item[4.] Fractional imputation (FI) method:  Use $y|(x, \delta=1) \sim N(\beta_0+\beta_1x_i, \sigma^2)$ and the response mechanism in  (\ref{eq:res_asumme}) to obtain the predictive model. The maximum likelihood estimator of $\theta$ is computed by using Fractional Imputation (FI) method in \citet{kim2011parametric}. Set the size of FI is 20. A description of the FI algorithm is described in Appendix D.
	
	\item[5.] Bayesian Data Augmentation (BDA)  method: Apply the proposed method in Section \ref{sec:NMAR} using the same model for FI method.   In the data augmentation algorithm, we choose the burn-in size as 2,000 and after burn-in, iteration size is 2,000.
\end{itemize}
Thus, in the last two methods, the outcome model is misspecified under $m_2$ and $m_3$. Under $\mathcal R_2$, the response mechanism is slightly  misspecified.

\begin{table}[ht]
	\centering
	\caption{Simulation results: ``m" denotes mean function,  ``bias" is the estimator subtracting true value,  ``R\_std" is the relative standard error which is relative to the standard error of full sample estimator. }\label{tbl:NMAR}
	\begin{tabular}{l|l|lrr|l|l|lrr}
		\hline
		Response  & m & method & bias & R\_std & Response & m & method & bias & R\_std \\ 
		\hline
		\multirow{15}{*}{$\mathcal R_1$} & \multirow{5}{*}{$m_1$} &  CC & -0.16 & 1.16  & \multirow{15}{*}{$\mathcal R_2$} & \multirow{5}{*}{$m_1$} & CC & -0.14 & 1.17\\ 
		&  & KC & -0.00 & 1.10 & & &   KC & -0.00 & 1.09 \\
		&  & FI & -0.00 & 1.09  & & & FI & -0.00 & 1.08 \\ 
		&  & BDA &  0.00 & 1.10  & & & BDA & -0.00 & 1.09 \\
		\cline{2-5}\cline{7-10}
		& \multirow{5}{*}{$m_2$} & CC & -0.18 & 1.14  &  & \multirow{5}{*}{$m_2$} & CC & -0.14 & 1.12\\ 
		&  &KC & 0.00 & 1.11  & &  & KC & -0.00 & 1.09 \\ 
		&  & FI & -0.01 & 1.10 & & & FI & -0.01 & 1.08  \\ 
		&  &  BDA & -0.00 & 1.11 & & & BDA & -0.00 & 1.09 \\
		\cline{2-5}\cline{7-10}
		& \multirow{5}{*}{$m_3$} &    CC & -1.13 & 1.15  &  & \multirow{5}{*}{$m_3$} &   CC & -0.95 & 1.13\\ 
		&  &KC & -0.00 & 1.03   & &  &  KC & 0.01 & 1.02\\ 
		&  & FI & -0.01 & 1.04   & &  & FI &0.01 & 1.03 \\ 
		&  & BDA & -0.00 & 1.04   & & & BDA & 0.01 & 1.03 \\
		\hline
	\end{tabular}
\end{table}

\begin{table}[ht]
	\centering
	\caption{The coverage probabilities for the proposed method}\label{tbl:cp_NMAR}
	\begin{tabular}{lllr}
		\hline
		method & m & res & cp \\ 
		\hline
		BDA & m1 & $\mathcal R_1$ & 0.95 \\ 
		BDA & m2 & $\mathcal R_1$ & 0.94 \\ 
		BDA & m3 & $\mathcal R_1$ & 0.95 \\ 
		\hline
		BDA & m1 & $\mathcal R_2$ & 0.95 \\ 
		BDA & m2 & $\mathcal R_2$ & 0.94 \\ 
		BDA & m3 & $\mathcal R_2$ & 0.95 \\ 
		\hline
	\end{tabular}
\end{table}

\newpage

The simulation results are presented in   Table \ref{tbl:NMAR} and \ref{tbl:cp_NMAR}. From Table \ref{tbl:NMAR}, we can see that the performance of the proposed BDA method is similar to the KC and FI methods. Furthermore, the proposed BDA method can simultaneously construct correct confidence intervals and does not involve Taylor linearization.  From Table \ref{tbl:cp_NMAR}, we can see that the coverage probabilities of the proposed method are around 0.95, which confirms the validity of the proposed BDA method.
\section{Application}\label{sec:application}

 
In this section, we apply the proposed Bayesian propensity score methods to Korea Labor and Income Panel Survey (KLIPS) data.
A brief description of the panel survey can be found at http://
www.kli.re.kr/klips/en/about/introduce.jsp.  The
study variable (y) is the average monthly income for the current
year and the auxiliary variable (x) can be demographic variables, such as the age groups and sex. 
Let $(X_i, Y_{it})$ be the observations for household $i$ in panel year $t$. The KLIPS  has $n=5,013$ households and $T=8$ panel years. We treat the first panel observations as the baseline measurements, and there are no missing data in the first year. In the panel survey, $X_i$ are completely observed and $Y_{it}$ are subject to missingness, for $i=1,2,\cdots, n$ and $t=1,2,\cdots, T$.  Let $\delta_{it}$ be the response indicator function of $Y_{it}$. Define
\begin{eqnarray}
\delta_{it}=\left\lbrace \begin{array}{ll}
1 & \text{if we observe $Y_{it}$}\\
0 & \text{otherwise.}
\end{array}\right. \nonumber
\end{eqnarray}
We are interested in estimating the probability of full response
\begin{eqnarray}
\pi_i=Pr(\delta_{i1}=1, \cdots, \delta_{iT}=1|X_i, Y_{i, obs}),\label{eq:pi_i}
\end{eqnarray}
where $Y_{i, obs}=(Y_{i1},\cdots, Y_{iT})'$ represents the observed responses for household $i$. The inverse of the $\pi_i$ in (\ref{eq:pi_i}) can be used as the propensity weight for the penal survey.  For monotone missing data, in the sense of $\delta_{it}=1$ implying $\delta_{i, t-1}=1, \cdots, \delta_{i1}=1$,  the probability reduces to 
\begin{eqnarray}
\pi_i=\pi_{i1}\pi_{i2}\cdots\pi_{iT}, \nonumber
\end{eqnarray} 
where $\pi_{it}=Pr(\delta_{it}=1|\delta_{i, t-1}=1, X_i, Y_{i1}, \cdots, Y_{i,t-1})$ under MAR assumption. 

For arbitrary missing patterns as in KLIPS, we first define $\delta_{it}^*=\prod_{k=1}^{t}\delta_{ik}$. Note that $\delta_{it}^*=1$ implies that $\delta_{i, t-1}^*=1$. Furthermore, 
\begin{eqnarray}
& Pr(\delta_{i1}=1, \cdots, \delta_{iT} &=1|X_i, Y_{i, obs})=Pr(\delta_{i1}^*=1, \cdots, \delta_{iT}^*=1|X_i, Y_{i, obs})\nonumber\\
&&= \prod_{k=2}^{T}Pr(\delta_{ik}^*=1|\delta_{i, k-1}^*=1, X_i, Y_{i, k-1})\nonumber\\
&&= \prod_{k=2}^{T}Pr(\delta_{ik}=1|\delta_{i, k-1}^*=1, X_i, Y_{i, k-1})\nonumber\\
&&= \pi_{i2}\pi_{i3}\cdots\pi_{iT}=\pi_i\nonumber,
\end{eqnarray}
where $\pi_{i1}=1$ for all samples.

Thus, we can build a parametric model for $\pi_{it}=Pr(\delta_{it}=1|\delta_{i, t-1}^*=1, X_i, Y_{i, t-1})$ and estimate the parameters sequentially.  Instead of using the frequentist approach of  \cite{zhou2012efficient}, we apply the BPS method in Section \ref{sec:proposed} and OBPS method in Section \ref{sec:extension} to incorporate the extra information in $X$.

We are interested in estimating the average income for the final year and constructing confidence intervals for the parameters.   Assume the response mechanism follows 
\begin{eqnarray}
\pi(\phi_t;X_i, Y_{i, t-1})=:Pr(\delta_{it}=1|\delta_{i, t-1}^*=1, X_i, Y_{i,t-1})=\frac{1}{1+\exp\left\lbrace -(X_i',Y_{i, t-1})\phi_t\right\rbrace }, \label{eq:phit}
\end{eqnarray}
which is known up to parameter $\phi_t$. Thus, we allow that the response probability at year $t$ depends on the last year income $y_{t-1}$, but not on the current year income. 
Assume $\delta_{it}$, given $\delta_{i, t-1}^*=1, X_i$, and $Y_{i, t-1}$, independently follow Bernoulli distribution with probability $\pi(\phi_t;X_i, Y_{i, t-1})$ in (\ref{eq:phit}). Therefore, the score function of $\phi_t$ is 
\begin{eqnarray}
S(\phi_t)=\frac{1}{n}\sum_{i=1}^{n}\left\lbrace \delta_{it}-\pi(\phi_t;X_i, Y_{i, t-1})\right\rbrace (X_i',Y_{i,t-1})' \delta_{i, t-1}^*.\nonumber
\end{eqnarray}

Then the joint estimating equations are $U_n(\phi_2,\phi_3,\cdots, \phi_T,\theta)=0$, where 
\begin{eqnarray}
U_n(\phi_2,\phi_3,\cdots, \phi_T,\theta)=n^{-1}\sum_{i=1}^{n}\left[ 
\begin{array}{l}
\left\lbrace \delta_{i2}-\pi(\phi_2;X_i, Y_{i,1})\right\rbrace (X_i',Y_{i,1})' \delta_{i, 1}^*\\
\vdots\\
\left\lbrace \delta_{iT}-\pi(\phi_T;X_i, Y_{i, T-1})\right\rbrace (X_i',Y_{i,T-1})' \delta_{i, T-1}^*\\
\pi_i^{-1}\delta^*_{iT} y_{i,T} - \theta  ,
\end{array}\right] \label{eq:app_joint_eq}
\end{eqnarray}
and $\theta=E(Y_T)$. 

The Bayesian propensity score (BPS) method can be described as 
\begin{itemize}
	\item [1. ] Solve $U_n(\phi_2,\phi_3,\cdots, \phi_T, \theta)=\mathbf 0$ to obtain $\hat \phi_2, \cdots, \hat \phi_T$, and $\hat \theta$.
	\item [2. ] Generate $\eta^*=(\eta_1^{*'} , \eta_2^{*'})'$ from $N (\mathbf 0, \hat \Sigma/n)$, where $\hat\Sigma$ is a consistent variance estimator of $\sqrt n U_n(\phi_2,\phi_3,\cdots, \phi_T, \theta) $. 
	\item [3. ] Solve $\left( S'(\phi_2), \cdots, S'(\phi_T)\right)'=\eta_1^* $ to obtain $\phi_2^*, \cdots, \phi_T^*$.
	\item [4. ] Compute $\pi_i^*=\pi(\phi_2^*;X_i, Y_{i, 1}) \times \cdots \times \pi(\phi_T^*;X_i, Y_{i, T-1})$. Solve 
	\begin{eqnarray}
	\frac{1}{n}\sum_{i=1}^{n}\frac{\delta_{iT}^*}{\pi_i^*} (y_{i,T} - \theta) =\eta_2^*\nonumber
	\end{eqnarray}
	to obtain $\theta^*$.
\end{itemize}
Repeat the above steps independently to generate samples from the posterior distribution of parameters. The variance-covariance matrix $\hat \Sigma$ can be derived by 
\begin{eqnarray}
\frac{1}{n}\sum_{i=1}^{n}\left[\begin{matrix}
\left\lbrace \delta_{i2}-\pi(\hat\phi_2;X_i, Y_{i,1})\right\rbrace (X_i',Y_{i,1})' \delta_{i, 1}^*\\
\vdots\\
\left\lbrace \delta_{iT}-\pi(\hat\phi_T;X_i, Y_{i, T-1})\right\rbrace (X_i',Y_{i,T-1})' \delta_{i, T-1}^*\\
\hat \pi_i^{-1}\delta^*_{iT} y_{i,T} - \hat{\theta}
\end{matrix}\right]^{\otimes 2}.\nonumber
\end{eqnarray}

To improve the efficiency of the point estimator, we also apply OBPS method to the same data. In addition to equations in (\ref{eq:app_joint_eq}), we add
\begin{eqnarray}
\sum_{i=1}^{n}\frac{\delta_{iT}^*}{\pi_i}\left( X_i-\mu_x\right) =0\nonumber\\
\sum_{i=1}^{n}\left(X_i-\mu_x \right) =0,\nonumber
\end{eqnarray}
where $\mu_x$ is the marginal proportion vector for demographical covariates. 
Therefore, the posterior distribution of $\theta$ can be obtained by applying the proposed algorithm in Section 5. 

%

For a comparison, we also considered a naive method which does not use the propensity model and apply the Bayesian method in the complete cases (CC) only. 
We apply BPS, OBPS and CC method to $T=2, 3,4$. The numerical results are presented below.


\begin{figure}[ht]
	\centering
	\includegraphics[width=0.9\textwidth]{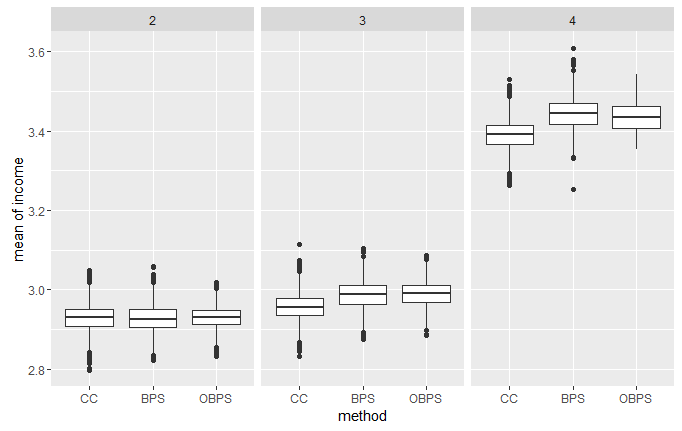}
	\caption{Boxplots for posterior distribution of $\theta$ by different methods and different panels. (Magnitude 1,000,000 Won)}\label{fig:income}
\end{figure}
From Figure \ref{fig:income}, all three methods provide similar estimators for the average income $\theta$.The trend of average income goes up as year $T$ increases. For year $T=2$, all three methods provide similar mean estimates. But the OBPS method is the most efficient. For year $T=3$, we see that the CC method provides lower mean estimate than BPS or OBPS, which is due to the nonresponse bias in the CC method. This phenomenon becomes more obvious for year $T=4$.  Also, the lengths of confidence intervals increase as $T$ increases, since the fully observed sample size is decreasing due to panel attrition. The CC method presents smaller values of $\theta$ for $T=4$, which suggests more panel attrition for higher income households. 
  Both BPS and OBPS provide similar mean estimates. But the OBPS method has narrower confidence intervals, which confirms the efficiency of the OBPS method.

\section{Concluding Remarks}\label{sec:Discussion}

A new Bayesian inference using PS method is developed using the idea of Approximate Bayesian computation. The proposed method can be widely applicable due to popularity of PS method. The proposed Bayesian approach is calibrated to frequentist inference in the sense that the proposed method provides the same inferential results with its frequentist version asymptotically \citep{little2012calibrated}. 
The calibration property holds if the prior distribution for the model parameters is flat. If the prior is informative then the resulting Bayesian inference   will be more efficient than frequentist inference thanks to its natural incorporation of the prior information. Thus, the proposed method is applicable when the need of combining information from different sources.

Causal inference, including estimation of average treatment effect from observational studies,  can be one promising application area of the PS method  (\citealp{morgan2014counterfactuals} and \citealp{hudgens2008toward}).  Developing tools for causal inference using the Bayesian PS method will be an important extension of this research.   Also,  Bayesian model selection method \citep{ishwaran2005spike} can be naturally applied to this setup. Such extensions will be topics for future research. 

\newpage

\appendix

\clearpage

\renewcommand{\theequation}{A.\arabic{equation}}
\renewcommand{\thesection}{A}
 \setcounter{equation}{0}

\section*{Appendix} 
\subsection*{A. Consistent variance estimator} \label{App:AppendixA}

From the asymptotic distribution in (\ref{asym2}), we can write
\begin{eqnarray}
\left[ U_n|\bm{\eta}\right] \sim N\left(\bm \eta, \Sigma/n \right) \nonumber.
\end{eqnarray}
To emphasize that $\Sigma$ is a function of $\phi, \theta$, we use $\Sigma=:\Sigma\left( \phi, \theta\right) $. Since the transformation
\begin{eqnarray}
\left( \begin{matrix}
\phi\\
\theta
\end{matrix}\right)\xrightarrow{} \left( \begin{matrix}
\bm \eta_1\\
\eta_2
\end{matrix}\right) \nonumber
\end{eqnarray} 
is one-to-one, $\Sigma\left( \phi, \theta\right)$ is equivalent to $\Sigma\left( \bm \eta_1, \eta_2\right)$, where $\boldsymbol{\eta}=(\boldsymbol{\eta}'_1, \eta_2)'$.
The corresponding density function is 
\begin{eqnarray}
p\left(U_n|\bm \eta \right)\propto \left| \Sigma\left(\bm \eta \right) /n\right|^{-1/2} \exp\left\lbrace -\frac{1}{2}\left(U_n-\bm \eta \right)'\left(\Sigma\left(\bm \eta \right)/n  \right)^{-1} \left(U_n-\bm \eta \right) \right\rbrace. \nonumber
\end{eqnarray}
Since we have assigned a flat prior, in the sense of $\pi\left(\bm \eta \right)\propto 1 $, we can derive the posterior distribution as
\begin{eqnarray}
p\left(\bm \eta|U_n \right) \propto \left| \Sigma\left(\bm \eta \right) /n\right|^{-1/2} \exp\left\lbrace -\frac{1}{2}\left(U_n-\bm \eta \right)'\left(\Sigma\left(\bm \eta \right)/n  \right)^{-1} \left(U_n-\bm \eta \right) \right\rbrace \nonumber.
\end{eqnarray}
By the definition of $\boldsymbol{\eta}$, $U_n$ is the unbiased estimator of $\boldsymbol{\eta}$. Thus, we write $ \boldsymbol{\hat \eta}=U_n$.
To show that 
\begin{eqnarray}
p\left(\bm \eta|U_n \right) \propto \left| \Sigma\left(\boldsymbol{\hat \eta} \right) /n\right|^{-1/2} \exp\left\lbrace -\frac{1}{2}\left(U_n-\bm \eta \right)'\left(\Sigma\left(\boldsymbol{\hat \eta} \right)/n  \right)^{-1} \left(U_n-\bm \eta \right) \right\rbrace ,  \nonumber 
\end{eqnarray}
 we first show that $\Sigma(\cdot)$ is continuous, which can be proved by the dominated convergence theorem applied to  $\bm\eta_1(\cdot)$ and $\eta_2(\cdot)$.  Now,  noting that, by asymptotic distribution (\ref{asym2}) and Chebyshev's inequality, we can show that $U_n\xrightarrow{P}\bm{\eta}$. Thus, we can obtain $\Sigma(\boldsymbol{\hat \eta})\xrightarrow{P} U_n(\bm \eta)$. Since $\Sigma$ is positive definite and $x^{-1/2}$ is continuous if $x>0$,  $|\Sigma(\boldsymbol{\hat \eta})|^{-1/2}\xrightarrow{P} |\Sigma(\bm \eta)|^{-1/2}$. By the continuous mapping theorem, 
\begin{eqnarray}
\sqrt n \Sigma(\boldsymbol{\hat \eta})^{-1/2}\left(U_n-\bm \eta \right)\xrightarrow{d} N(\mathbf 0, \mathbf I).\nonumber
\end{eqnarray}
Therefore, we can derive the posterior distribution as
\begin{eqnarray}
p\left(\bm \eta|U_n\right) \propto \left| \Sigma\left(\boldsymbol{\hat \eta}\right) /n\right|^{-1/2} \exp\left\lbrace -\frac{1}{2}\left(U_n-\bm \eta \right)'\left(\Sigma\left(\boldsymbol{\hat \eta}\right)/n  \right)^{-1} \left(U_n-\bm \eta \right) \right\rbrace \nonumber.
\end{eqnarray}

That is $[\bm\eta|U_n=\mathbf 0]\sim N(\mathbf 0, \Sigma(U_n=\mathbf 0)/n)$, which is equivalent to
\begin{eqnarray}
p\left(\bm\eta|U_n =\bm 0 \right) \propto \left| \Sigma(\hat \phi, \hat \theta) /n\right|^{-1/2} \exp\left[  -\frac{1}{2}\left(U_n-\bm \eta \right)'\left\lbrace \Sigma \left( \hat \phi, \hat \theta\right)/n  \right\rbrace ^{-1} \left(U_n-\bm \eta \right) \right]  \nonumber,
\end{eqnarray}
where $(\hat \phi, \hat \theta)$ is the solution to $U_n=\mathbf 0$.
Furthermore, the consistency of $\hat \Sigma=\hat \Sigma(\hat \phi, \hat \theta)$ in (\ref{eq:hat_sigma}) can be proved using the law of large numbers.

\renewcommand{\theequation}{B.\arabic{equation}}
\renewcommand{\thesection}{B}
 \setcounter{equation}{0} 
 
\subsection*{B. Proof of Theorem \ref{theorem3}} \label{App:AppendixB}
	\subsection*{Step I}
		
	From Condition [C9], we assume that $\bm\zeta\mapsto U_n(\bm\zeta) $ and $\bm\zeta\mapsto \bm \eta(\bm\zeta) $ are one-to-one functions, for any $\bm\zeta\in N( \bm\zeta_0) $. 	 Denote these two mappings as $T_n$ and $T$ respectively.
	Because of their one-to-one property, their inverse mappings exist for $\bm\zeta \in N_n( \bm\zeta_0) $. Therefore, we can write (\ref{12}) as
	\begin{eqnarray}
	\sqrt n\left( U_n-\bm \eta\right)\xrightarrow{d} N[0, \Sigma\left\lbrace  T^{-1}\left(\bm \eta \right) \right\rbrace  ] \nonumber,
	\end{eqnarray}
	which leads to
	\begin{eqnarray}
	p(U_n|\bm \eta ) \xrightarrow{}\phi_{\bm \eta,n^{-1} \Sigma\left( T^{-1}\left(\bm \eta \right) \right)}\left(U_n \right) \label{asy1}\nonumber.
	\end{eqnarray}
	Thus, by the convergence of $U_n$ to $\bm\eta$ and 
	using the argument similar to the proof for Lemma 1 in \cite{soubeyrand2015weak}, we can show that
	\begin{eqnarray}
	p(\bm \eta|U_n) =\phi_{U_n,n^{-1}\Sigma\left(T_n^{-1} \left(U_n \right) \right) }( \bm \eta)\left\lbrace  1+o_p(1)\right\rbrace  \label{sym3}.
	\end{eqnarray}

	\subsection*{Step II}
	Note that $U_n(\hat{\bm\zeta})=0$, thus $T_n^{-1}(0)=\hat{\bm\zeta}$. 
	From (\ref{sym3}), we can therefore get the posterior distribution
	\begin{eqnarray}
	p(\bm \eta|U_n=0 )=p(\bm \eta|\hat{\bm\zeta})=\phi_{0,n^{-1}\Sigma( \hat{\bm\zeta} )}( \bm \eta) \left\lbrace  1+o_p(1)\right\rbrace .\label{claim_p}
	\end{eqnarray}
    Thus,  we can write the density $p(\bm \eta|U_n=0 )$ as 
	\begin{eqnarray}
	\phi_{0,n^{-1}\Sigma( \hat{\bm\zeta} )}( \bm \eta)\propto \exp\left\lbrace-\frac{n}{2} \bm \eta'\Sigma^{-1}( \hat{\bm\zeta})\bm \eta  \right\rbrace \nonumber.
	\end{eqnarray}
	Furthermore, by the consistency of the variance estimator provided in condition [C7], we can obtain $\hat \Sigma:=\hat \Sigma(\hat{ \bm{\zeta}})=\Sigma(\hat {\bm\zeta})\left\lbrace 1+o_p(1)\right\rbrace  $. 
	Thus, 
	\begin{eqnarray}
	\bm \eta'\Sigma^{-1}( \hat{\bm\zeta})\bm \eta =\bm \eta'\left\lbrace {\hat  \Sigma}^{-1}( 1+o_p(1 ) )  \right\rbrace \bm \eta= \bm \eta'\bm {\hat \Sigma}^{-1} \bm \eta\left\lbrace 1 +o_p(1)\right\rbrace     ,\nonumber
	\end{eqnarray}
	which leads to 
	\begin{eqnarray}
	\phi_{0,n^{-1}\Sigma( \hat{\bm\zeta} )}( \bm \eta)=\phi_{0,n^{-1} {\hat \Sigma} }( \bm \eta)\left\lbrace 1+o_p( 1)\right\rbrace \exp\left\lbrace -\frac{n}{2}o_p\left( \bm \eta'\bm {\hat \Sigma}^{-1} \bm \eta\right) \right\rbrace  \nonumber.
	\end{eqnarray}
	From [C1] and [C5], we have $U_n(\bm \zeta)\xrightarrow{}\bm \eta$ in probability and $U_n=O_p(1/\sqrt{n})$ for $\bm \zeta\in N(\bm \zeta_0)$, which leads to $\boldsymbol{\eta}=O(1/\sqrt n)$. 
	Thus, 
	\begin{eqnarray}
	\exp\left\lbrace -\frac{n}{2}o_p\left( \bm \eta'\bm {\hat \Sigma}^{-1} \bm \eta\right) \right\rbrace=\exp\left\lbrace o_p(1)\right\rbrace\xrightarrow{}1,\nonumber
	\end{eqnarray}
	in probability and the following follows
	\begin{eqnarray}
	p(\bm \eta|U_n=0 )=p(\bm \eta|\hat{\bm\zeta} )=\phi_{0,n^{-1} {\hat \Sigma} }( \bm \eta)\left\lbrace 1+o_p( 1)\right\rbrace  \label{asym4}.
	\end{eqnarray}
	
	\subsection*{Step III}
	Let $\bm\eta^*$ be generated from the asymptotic posterior distribution (\ref{asym4}) which is a normal distribution with mean 0 and variance $\hat\Sigma/n $. 
%
	Therefore, the $j$-th component $\bm\zeta_j^*$ of $\bm\zeta^*$ satisfies
	\begin{eqnarray}
	&E\left\lbrace \bm\zeta^*_j|\hat{\bm\zeta}_n\right\rbrace &=E\left\lbrace T_{n,j}^{-1} (\bm\eta^* )|\hat{\bm\zeta}_n \right\rbrace\nonumber\\
	&&=E\left\lbrace \left. T_{n,j}^{-1}\left( 0\right) +\left.\frac{\partial T_{n,j}^{-1}(\bm\eta ) }{\partial \bm\eta'}\right|_{\bm\eta=0}\eta^*+\frac{1}{2}{\bm\eta^*}'\left.\frac{\partial^2T_{n,j}^{-1}(\bm\eta )}{\partial \bm\eta\bm\eta'}\right|_{\bm\eta=0}\bm\eta^*+o_p({\bm\eta^*}'\bm\eta^*)\right|\hat{\bm\zeta}_n\right\rbrace \nonumber\\
	&&=\hat{\bm\zeta}_{n,j}+\frac{1}{2}E\left\lbrace\left.{\bm\eta^*}'\left.\frac{\partial^2T_{n,j}^{-1}(\bm\eta )}{\partial \bm\eta\bm\eta'}\right|_{\bm\eta=0}\bm\eta^*\right|\hat{\bm\zeta}_n \right\rbrace+ o\left(\frac{1}{n} \right) \nonumber.
	\end{eqnarray}
	
	By $ E(\bm Z'\Lambda \bm Z )=tr(\Lambda\Sigma )+\mu'\Lambda\mu   $, we derive
	\begin{eqnarray}
	E\left\lbrace\left.{\bm\eta^*}'\left.\frac{\partial^2T_{n,j}^{-1}(\bm\eta )}{\partial \bm\eta\bm\eta'}\right|_{\bm\eta=0}\bm\eta^*\right|\hat{\bm\zeta}_n \right\rbrace=tr\left[\left.\frac{\partial^2T_{n,j}^{-1}(\bm\eta )}{\partial \bm\eta\bm\eta'}\right|_{\bm\eta=0}\frac{ {\hat \Sigma} }{n} \right] =O\left(\frac{1}{n} \right) \nonumber,
	\end{eqnarray}
	under [C4]. 
	Therefore, we have 
	\begin{eqnarray}
	E\left\lbrace \bm\zeta^*_j|\hat{\bm\zeta}\right\rbrace=\hat{\bm\zeta}_{n,j}+O\left( \frac{1}{n}\right)  \nonumber,
	\end{eqnarray}
	for $j=1,2,\cdots, p$,
	which establishes 
	\begin{eqnarray}
	E\left\lbrace \bm\zeta^*|\hat{\bm\zeta}\right\rbrace=\hat{\bm\zeta}_{n}+O\left( \frac{1}{n}\right) \label{III}.
	\end{eqnarray}
	
	\subsection*{Step IV}
	Now, the posterior variance of $\bm\zeta^*_j$:
	\begin{eqnarray}
	&Var\left\lbrace \bm\zeta^*_j|\hat{\bm\zeta}\right\rbrace&=Var\left\lbrace \left. T_{n,j}^{-1}\left( 0\right) +\left.\frac{\partial T_{n,j}^{-1}(\bm\eta) }{\partial \bm\eta'}\right|_{\bm\eta=0}\bm\eta^*+\frac{1}{2}{\bm\eta^*}'\left.\frac{\partial^2T_{n,j}^{-1}(\bm\eta )}{\partial \bm\eta\bm\eta'}\right|_{\bm\eta=0}\bm\eta^*+o_p({\bm\eta^*}'\bm\eta^*)\right|\hat{\bm\zeta}\right\rbrace \nonumber \\
	&&=Var\left\lbrace \left. \left.\frac{\partial T_{n,j}^{-1}(\bm\eta ) }{\partial \bm\eta'}\right|_{\bm\eta=0}\bm\eta^*+\frac{1}{2}{\bm\eta^*}'\left.\frac{\partial^2T_{n,j}^{-1}(\bm\eta)}{\partial \bm\eta\bm\eta'}\right|_{\bm\eta=0}\bm\eta^*+o_p({\bm\eta^*}'\bm\eta^*)\right|\hat{\bm\zeta}\right\rbrace \nonumber.
	\end{eqnarray}
	The first term is
	\begin{eqnarray}
	&Var\left\lbrace  \left. \left.\frac{\partial T_{n,j}^{-1}(\bm\eta) }{\partial \bm\eta'}\right|_{\bm\eta=0}\bm\eta^*\right|\hat{\bm\zeta}\right\rbrace&=\left.\frac{\partial T_{n,j}^{-1}(\bm\eta ) }{\partial \bm\eta'}\right|_{\bm\eta=0} Var\left\lbrace \bm\eta^*|\hat{\bm\zeta}\right\rbrace \left\lbrace \left.\frac{\partial T_{n,j}^{-1}(\bm\eta ) }{\partial \bm\eta'}\right|_{\bm\eta=0}\right\rbrace'\nonumber\\
	&&=O\left( \frac{1}{n}\right). \label{I1}
	\end{eqnarray}
	For the second term, using
	\begin{eqnarray}
	Var(\bm Z' \Lambda \bm Z )= 2 tr(\Lambda \Sigma \Lambda\Sigma ) +4\mu'\Lambda\Sigma\Lambda \nonumber,\\
	Cov(\bm Z' \Lambda_1 \bm Z, \bm Z' \Lambda_2 \bm Z ) =2 tr(\Lambda_1 \Sigma \Lambda_2\Sigma ) +4\mu'\Lambda_1\Sigma\Lambda_2 \nonumber,
	\end{eqnarray}
	for $\bm Z \sim N( \mu ,\Sigma) $, 
    we have
	\begin{eqnarray}
	Var\left\lbrace \left. {\bm\eta^*}'\left.\frac{\partial^2T_{n,j}^{-1}(\bm\eta )}{\partial \bm\eta\bm\eta'}\right|_{\bm\eta=0}\bm\eta^*\right|\hat{\bm\zeta}\right\rbrace=2tr\left\lbrace \left.\frac{\partial^2T_{n,j}^{-1}(\bm\eta )}{\partial \bm\eta\bm\eta'}\right|_{\bm\eta=0}\frac{{\hat \Sigma} }{n} \left.\frac{\partial^2T_{n,j}^{-1}(\bm\eta)}{\partial \bm\eta\bm\eta'}\right|_{\bm\eta=0}\frac{{\hat \Sigma} }{n}\right\rbrace\nonumber =O\left(\frac{1}{n^2} \right).
	\end{eqnarray}
	 The covariance of two terms is less than the square root of their variances. We have shown that the variance of the first term is in the order of $O( 1/n) $ and the variance of the second term is in the order of $O( 1/n^2) $. So the covariance has  the order of $O( n^{-3/2}) $.
	
	Similarly, we can derive
	\begin{eqnarray}
	Cov(\bm\zeta^*_j,\bm\zeta^*_k|\hat{\bm\zeta} )=\left.\frac{\partial T_{n,j}^{-1}(\bm\eta ) }{\partial \bm\eta'}\right|_{\bm\eta=0} Var\left\lbrace\bm\eta^*|\hat{\bm\zeta} \right\rbrace \left\lbrace \left.\frac{\partial T_{n,k}^{-1}(\bm\eta) }{\partial \bm\eta'}\right|_{\bm\eta=0}\right\rbrace'+o\left(\frac{1}{n}\right) . \label{I2}
	\end{eqnarray}
	Combining (\ref{I1}) and (\ref{I2}), we have 
	\begin{eqnarray}
	Var(\zeta^*|\hat{\bm\zeta}) = \left.\frac{\partial T_{n}^{-1}(\bm\eta) }{\partial \bm\eta'}\right|_{\bm\eta=0} Var\left\lbrace \bm\eta^*|\hat{\bm\zeta} \right\rbrace \left\lbrace\left.\frac{\partial T_{n}^{-1}(\bm\eta) }{\partial \bm\eta'}\right|_{\bm\eta=0} \right\rbrace'+o\left(\frac{1}{n}\right) . \label{IV}
	\end{eqnarray}
	\subsection*{Step V}

	By Conditions [C1]-[C5], we have 
	\begin{eqnarray}
	\sqrt n (\hat{\bm\zeta}-\bm\zeta_0 ) \xrightarrow{d} N(0, A^{-1}( \bm\zeta_0) \Sigma( \bm\zeta_0) A'^{-1}( \bm\zeta_0) ), \label{asymp5}\nonumber
	\end{eqnarray}
	where $A( \bm\zeta) = \partial \boldsymbol{\eta}(\bm \zeta)/\partial \bm\zeta$. See Theorem 5.21 in \cite{van2000asymptotic}.
	
	Since $T_n\xrightarrow{}T$ uniformly by [C1] and both mappings are one-to-one functions, we can state that $T^{-1}_n\xrightarrow{}T^{-1}$ uniformly. Thus,
	\begin{eqnarray}
	\left.\frac{\partial T_{n}^{-1}(\bm\eta) }{\partial \bm\eta'}\right|_{\bm\eta=0}\xrightarrow{P}\left.\frac{\partial T^{-1}(\bm\eta) }{\partial \bm\eta'}\right|_{\bm\eta=0}=A^{-1}(\bm \zeta_0).\nonumber 
	\end{eqnarray}
	Also, by [C7], we have ${\hat \Sigma}\xrightarrow{P} \Sigma(\hat{\bm\zeta})  $ and $\hat{\bm\zeta}\xrightarrow{P}\bm\zeta_0$. By the Lipschitz continuity of $\Sigma(\zeta)$ , we can conclude that $ \Sigma(\hat{\bm\zeta})\xrightarrow{P}\Sigma(\bm\zeta_0) $. Thus, $\hat\Sigma\xrightarrow{P} \Sigma(\bm\zeta_0)$
	
	and
	\begin{eqnarray}
	&nVar( \bm\zeta^*|\hat{\bm\zeta})-nVar(\hat{\bm\zeta} )&= \left.\frac{\partial T_{n}^{-1}(\bm\eta) }{\partial \bm\eta'}\right|_{\bm\eta=0} {\hat \Sigma}  \left\lbrace\left.\frac{\partial T_{n}^{-1}(\bm\eta) }{\partial \bm\eta'}\right|_{\bm\eta=0}\right\rbrace'\nonumber\\
	&&- A^{-1}( \bm\zeta_0) \Sigma(\bm\zeta_0) A'^{-1}( \bm\zeta_0) \xrightarrow{P}0\label{claim4},
	\end{eqnarray}
	by the continuous mapping theorem. 
	Combining the previous conclusions (\ref{III}), (\ref{IV}) and (\ref{claim4}), we can use Slutsky's theorem to get
	\begin{eqnarray}
	\left\lbrace Var(\hat{\bm\zeta} )\right\rbrace^{-1/2}( \bm\zeta^*-\hat{\bm\zeta}) |\hat{\bm\zeta}\xrightarrow{d} N(0,\bm I_p ) \label{asymp6},\nonumber
	\end{eqnarray}
	which proves (\ref{21}). 
	
	\subsection*{ Step VI } 
	Let $\alpha\in(0,1 ) $, and define 
\begin{eqnarray}
C_{n,\alpha}=\left\lbrace \bm\zeta^*:(\hat {\bm\zeta}-\bm\zeta^* )'\left\lbrace Var(\hat{\bm\zeta} ) \right\rbrace^{-1}(\hat {\bm\zeta}-\bm\zeta^* )\leq \chi_p^2(\alpha)    \right\rbrace \nonumber,
\end{eqnarray}
where the $\chi_p^2(\alpha) $ is the $\alpha$ quantile of Chi-square distribution with $p$ degrees of freedom. 


Furthermore, from a property of the Raylei quotient \citep{horn1985matrix}, there exists a matrix $O$ such that 
\begin{eqnarray}
O\left\lbrace  Var^{-1}(\hat{\bm\zeta})/n\right\rbrace O^T=\text{diag}\left\lbrace \lambda_1, \cdots, \lambda_p \right\rbrace , \nonumber
\end{eqnarray}
where $OO^T=\bm I_p$ and $0<\lambda_1\leq \lambda_2,\cdots, \leq \lambda_p$. 
Thus we obtain 
\begin{eqnarray}
\bm x^T\left\lbrace  Var^{-1}(\hat{\bm\zeta} )\right\rbrace\bm x\geq n\lambda_1\bm x^T\bm x\label{eq:ineq1}.
\end{eqnarray}

Also, we can apply the conclusion (\ref{eq:ineq1}) to get 
\begin{eqnarray}
\|\hat {\bm\zeta}-\bm\zeta^* \|\leq \lambda_1^{-1/2} \sqrt{(\hat{\bm \zeta}-\bm\zeta^* )'\left\lbrace Var(\hat{\bm\zeta}) \right\rbrace^{-1}(\hat{\bm \zeta}-\bm\zeta^* )/n}\leq \lambda_1^{-1/2}\sqrt{\chi_p^2(\alpha)/n }\nonumber 
\end{eqnarray}
 for all $\bm\zeta^* \in C_{n, \alpha}$. 
 
 Similarly, by the asymptotic normality of the estimator $\hat {\bm\zeta}$ and applying the conclusion (\ref{eq:ineq1}),
 \begin{eqnarray}
 \| \hat {\bm\zeta}-\bm\zeta_0\|\leq \lambda_1^{-1/2} \sqrt{( \hat{\bm \zeta}-\bm\zeta_0)^T\left\lbrace Var( \hat{\bm \zeta)} \right\rbrace^{-1} ( \hat {\bm\zeta}-\bm\zeta_0)}\leq \lambda_1^{-1/2}\sqrt{\chi_p^2(\alpha)/n }\label{eq:ineq_2}.
 \end{eqnarray}
 Next, from (\ref{eq:ineq_2}), we can conclude that 
 \begin{eqnarray}
\lim_{n\xrightarrow{}\infty} Pr\left(  \| \hat{\bm \zeta}-\bm\zeta_0\|\leq\lambda_1^{-1/2}\sqrt{\chi_p^2(\alpha )/n }\right) \geq \alpha\nonumber.
 \end{eqnarray}
 By the inequality $\|\bm\zeta^*-\bm\zeta_0 \|\leq\|\hat {\bm\zeta}-\bm\zeta^* \|+\| \hat{\bm \zeta}-\bm\zeta_0\|  $, we obtain 
 \begin{eqnarray}
 \lim_{n\xrightarrow{}\infty} Pr\left(\forall \bm\zeta^* \in C_{n, \alpha}, \quad \left\|\bm\zeta^*-\bm\zeta_0 \right\|\leq 2\lambda_1^{-1/2}\sqrt{\chi_p^2(\alpha)/n }\right)\geq \alpha \label{eq:ineq_one}.
 \end{eqnarray}
 Since we have defined $N_n(\bm\zeta_0) $ in a neighborhood with center $\bm\zeta_0$ and radius $r_n$, where $r_n$ satisfies $r_n\xrightarrow{}0$ and $\sqrt{n} r_n\xrightarrow{}\infty$. From (\ref{eq:ineq_one}),
 \begin{eqnarray}
  \lim_{n\xrightarrow{}\infty} Pr(\forall \bm\zeta^* \in C_{n, \alpha}, \quad \left\|\bm\zeta^*-\bm\zeta_0 \right\|\leq r_n)\geq \alpha\nonumber,\\
\lim_{n\xrightarrow{}\infty} Pr( C_{n,\alpha}\subset N_n(\bm\zeta_0 ) ) \geq \alpha\nonumber.
 \end{eqnarray}
 Therefore, 
 \begin{eqnarray}
  \lim_{n\xrightarrow{}\infty} Pr\left(\int_{N_n(\bm\zeta_0) }\phi_{\hat{\bm\zeta},Var(\hat{\bm\zeta})}( \bm\zeta^*) d\bm\zeta^*\geq \int_{C_{n,\alpha}}\phi_{\hat{\bm\zeta},Var(\hat{\bm\zeta})}( \bm\zeta^*) d\bm\zeta^* \right) \geq \alpha.\nonumber
 \end{eqnarray}
 This is equivalent to 
 \begin{eqnarray}
  \lim_{n\xrightarrow{}\infty} Pr\left(\int_{N_n(\bm\zeta_0) }\phi_{\hat{\bm\zeta},Var(\hat{\bm\zeta} )}( \bm\zeta^*) d\bm\zeta^*\geq \alpha \right) \geq \alpha\nonumber.
 \end{eqnarray}
 The above conclusion holds for any $\alpha\in( 0,1) $. Thus
 \begin{eqnarray}
 \lim_{n\xrightarrow{}\infty} \int_{N_n(\bm\zeta_0 )} \phi_{\hat{\bm\zeta},Var(\hat{\bm\zeta})}( \bm\zeta^*) d\bm\zeta^* =1 \quad \text{in probability}.\nonumber
 \end{eqnarray}

\renewcommand{\theequation}{C.\arabic{equation}}
\renewcommand{\thesection}{C}
 \setcounter{equation}{0} 

\subsection*{C. Computational Details for the Metropolis-Hastings Algorithm } \label{App:AppendixD}

 Implementing the optimal Bayesian propensity score (OBPS) method is done through the following algorithm.
\begin{itemize}
	\item[1.] Choose the initial value for $\boldsymbol{\psi}$ and denote it as $\boldsymbol{\psi}_0$. 
	\item[2.] For iteration $t$, given the current parameter value $\boldsymbol{\psi}_t$, generate $\Delta\boldsymbol{\psi}$ from $N(\mathbf 0, V)$, where $V$ is a tunning parameter obtainable by the data-driven method discussed below.  Let the candidate value be $\boldsymbol{\psi}^*=\boldsymbol{\psi}_t+\Delta\boldsymbol{\psi}$.
	\item[3.] Compute the acceptance probability $$\alpha=\alpha(\boldsymbol{\psi}^*|\boldsymbol{\psi}_t) =\min\left\lbrace 1, \frac{g\left(U_n|\boldsymbol{\psi}^* \right)\pi(\boldsymbol{\psi}^*) }{g\left(U_n|\boldsymbol{\psi}_t \right)\pi(\boldsymbol{\psi}_t)} \right\rbrace. $$
	\item[4.] Generate $u$ from Uniform $(0,1)$ distribution. If $u<\alpha$, accept the candidate $\boldsymbol{\psi}_{t+1}=\boldsymbol{\psi}^*$. Otherwise let $\boldsymbol{\psi}_{t+1}=\boldsymbol{\psi}_t$.
	\item[5.] For burning in period, discard the values from the first $B$ iterations. Then collect $M$ values. These $M$ values can be treated as values generated from the target posterior distribution.
\end{itemize}	

For the choice of the initial value for $\boldsymbol{\psi}$, we can use the solution to $$(U_1'(\phi), U_2(\phi,\theta), U_4'(\bmu_x))'=\mathbf 0.$$ In Metropolis-Hastings algorithm, the value of $V$ for the random walk will directly affect the speed of convergence of the Markov chain and the acceptance rate. We recommend a data-driven method to set $V$. A data-driven choice of $V$ can be obtained from the posterior variance of the Monte Carlo samples from 
 $p\{\bmu_x, \phi, \theta|(U_1, U_2, U_4)=\mathbf 0\}$. 

To compute the acceptance probability, we need to compute the ratio
\begin{eqnarray}
\frac{g\left( U_n|\boldsymbol{\psi}^*\right) }{g\left( U_n|\boldsymbol{\psi}_t\right)}=\frac{|\Sigma(\boldsymbol{\psi}^*)|^{-1/2}\exp\left\lbrace -\frac{n}{2}U'_n(\boldsymbol{\psi}^*)\Sigma^{-1}(\boldsymbol{\psi}^*)U_n(\boldsymbol{\psi}^*)\right\rbrace }{|\Sigma(\boldsymbol{\psi}_t)|^{-1/2}\exp\left\lbrace -\frac{n}{2}U'_n(\boldsymbol{\psi}_t)\Sigma^{-1}(\boldsymbol{\psi}_t)U_n(\boldsymbol{\psi}_t)\right\rbrace}, \label{eq:ratio}\nonumber
\end{eqnarray}
which  can be approximated by 
\begin{eqnarray}
\exp\left\lbrace-\frac{n}{2}U'_n(\boldsymbol{\psi}*)\hat\Sigma^{-1}U_n(\boldsymbol{\psi}*) +\frac{n}{2} U'_n(\boldsymbol{\psi}_t)\hat\Sigma^{-1}U_n(\boldsymbol{\psi}_t) \right\rbrace ,\nonumber
\end{eqnarray}
where 
\begin{eqnarray}
\hat \Sigma=\frac{1}{n}\sum_{i=1}^{n}\left(\begin{matrix}
s(\hat \phi;\mathbf x_i)\\
\delta_i\hat\pi_i^{-1}U(\hat \theta; \mathbf x_i, y_i)\\
\delta_i\hat\pi_i^{-1}(\mathbf x_i -\hat \bmu_x)\\
(\mathbf x_i -\hat \bmu_x)
\end{matrix} \right)^{\otimes 2} ,\nonumber
\end{eqnarray} 
and $(\hat \bmu_x, \hat \phi,\hat \theta)$ are the consistent estimators.

\renewcommand{\theequation}{D.\arabic{equation}}
\renewcommand{\thesection}{D}
 \setcounter{equation}{0} 

\subsection*{D. Fractional imputation algorithm in simulation study two}\label{App:AppendixE}

Let $s(\phi;\delta_i, \mathbf{x}_1, y)$ be the score function for the response model. 
In additional to the response model, we also assume $f(y\mid x, \delta=1;\gamma)$.  To solve the observed score function of $\phi$, that is 
\begin{eqnarray}
\bar S(\phi)=\sum_{i=1}^{n}\left[\delta_is(\phi;\delta_i, x_i, y_i)+(1-\delta_i)E\left\lbrace s(\phi;\delta_i,x_i, y)|x_i, \delta_i=0\right\rbrace  \right] =0, \label{d-1}
\end{eqnarray}
where the conditional expectation is with respect to the prediction model in (\ref{eq:link}). The estimate $\hat \gamma$ of $\gamma$ can be obtained by using the observed data to fit the model $f(y\mid x, \delta=1;\gamma)$. To compute the solution $\hat{\phi}$ to (\ref{d-1}), EM algorithm using fractional imputation   \citep{kim2011parametric} can be used.   The algorithm is described as followings:
\begin{itemize}
	\item [\textbf{E-step}:] For each unit $i$ with $\delta_i=0$, generate $y_{ij}^*$ from $f_1(y\mid x_i, \delta_i=1;\hat \gamma)$ for $j=1,2,\cdots, b$.   Given the current value of $\phi_1$, compute the fractional weights of $y_{ij}$ as
	\begin{eqnarray}
	w_{ij}^*\propto O(\phi;y_{ij}^*)=\frac{1-Pr(\delta_i=1\mid x_i, y_{ij}^* )}{Pr(\delta_i=1\mid x_i, y_{ij}^* )}\propto\exp(-\phi_1y_{ij}^*),\nonumber
	\end{eqnarray}
	subject to $\sum_{j=1}^{b}w_{ij}^* =1$.
	\item [\textbf{M-step}:] Update $\phi$ by solving 
	\begin{eqnarray}
	\bar S(\phi)=\sum_{i=1}^{n}\left[\delta_is(\phi;\delta_i, x_i, y_i)+(1-\delta_i)\sum_{j=1}^{b}w_{ij}^* s(\phi;\delta_i,x_i, y_{ij}^*) \right] =0.\nonumber
	\end{eqnarray}
\end{itemize}
Repeat \textbf{E-step} and \textbf{M-step} iteratively until convergence.  After convergence, the final estimator of $\theta=E(Y)$ is constructed by 
\begin{eqnarray}
\hat \theta_{FI} = \frac{1}{n} \sum_{i=1}^{n} \left\{  \delta_iy_i +(1-\delta_i) \sum_{j=1}^b w_{ij}^*  y_{ij}^*\right\}.\nonumber
\end{eqnarray}

\newpage

\bibliographystyle{chicago}
\bibliography{ref}

\begin{thebibliography}{}

\bibitem[\protect\citeauthoryear{An}{An}{2010}]{an2010bayesian}
An, W. (2010).
\newblock {B}ayesian propensity score estimators: incorporating uncertainties
  in propensity scores into causal inference.
\newblock {\em Sociological Methodology\/}~{\em 40\/}(1), 151--189.

\bibitem[\protect\citeauthoryear{Bang and Robins}{Bang and
  Robins}{2005}]{bang2005doubly}
Bang, H. and J.~M. Robins (2005).
\newblock Doubly robust estimation in missing data and causal inference models.
\newblock {\em Biometrics\/}~{\em 61\/}(4), 962--973.

\bibitem[\protect\citeauthoryear{Beaumont, Zhang, and Balding}{Beaumont
  et~al.}{2002}]{beaumont2002approximate}
Beaumont, M.~A., W.~Zhang, and D.~J. Balding (2002).
\newblock Approximate {B}ayesian {C}omputation in population genetics.
\newblock {\em Genetics\/}~{\em 162\/}(4), 2025--2035.

\bibitem[\protect\citeauthoryear{Cao, Tsiatis, and Davidian}{Cao
  et~al.}{2009}]{cao2009improving}
Cao, W., A.~A. Tsiatis, and M.~Davidian (2009).
\newblock Improving efficiency and robustness of the doubly robust estimator
  for a population mean with incomplete data.
\newblock {\em Biometrika\/}~{\em 96\/}(3), 723--734.

\bibitem[\protect\citeauthoryear{Chen and Shao}{Chen and
  Shao}{1999}]{chen1999monte}
Chen, M.-H. and Q.-M. Shao (1999).
\newblock {M}onte {C}arlo estimation of {B}ayesian credible and {HPD}
  intervals.
\newblock {\em Journal of Computational and Graphical Statistics\/}~{\em
  8\/}(1), 69--92.

\bibitem[\protect\citeauthoryear{Chib and Greenberg}{Chib and
  Greenberg}{1995}]{chib1995understanding}
Chib, S. and E.~Greenberg (1995).
\newblock Understanding the {M}etropolis-{H}astings {A}lgorithm.
\newblock {\em The American Statistician\/}~{\em 49\/}(4), 327--335.

\bibitem[\protect\citeauthoryear{Flanders and Greenland}{Flanders and
  Greenland}{1991}]{flanders1991analytic}
Flanders, W.~D. and S.~Greenland (1991).
\newblock Analytic methods for two-stage case-control studies and other
  stratified designs.
\newblock {\em Statistics in Medicine\/}~{\em 10\/}(5), 739--747.

\bibitem[\protect\citeauthoryear{Horn and Johnson}{Horn and
  Johnson}{1985}]{horn1985matrix}
Horn, R.~A. and C.~R. Johnson (1985).
\newblock Matrix {A}nalysis {C}ambridge {U}niversity {P}ress.
\newblock {\em New York\/}.

\bibitem[\protect\citeauthoryear{Hudgens and Halloran}{Hudgens and
  Halloran}{2008}]{hudgens2008toward}
Hudgens, M.~G. and M.~E. Halloran (2008).
\newblock Toward causal inference with interference.
\newblock {\em Journal of the American Statistical Association\/}~{\em
  103\/}(482), 832--842.

\bibitem[\protect\citeauthoryear{Hyndman}{Hyndman}{1996}]{hyndman1996computing}
Hyndman, R.~J. (1996).
\newblock Computing and graphing highest density regions.
\newblock {\em The American Statistician\/}~{\em 50\/}(2), 120--126.

\bibitem[\protect\citeauthoryear{Imai and Ratkovic}{Imai and
  Ratkovic}{2014}]{imai2014covariate}
Imai, K. and M.~Ratkovic (2014).
\newblock Covariate balancing propensity score.
\newblock {\em Journal of the Royal Statistical Society: Series B (Statistical
  Methodology)\/}~{\em 76\/}(1), 243--263.

\bibitem[\protect\citeauthoryear{Ishwaran and Rao}{Ishwaran and
  Rao}{2005}]{ishwaran2005spike}
Ishwaran, H. and J.~S. Rao (2005).
\newblock Spike and {S}lab variable selection: frequentist and {B}ayesian
  strategies.
\newblock {\em Annals of Statistics\/}~{\em 33\/}(2), 730--773.

\bibitem[\protect\citeauthoryear{Kaplan and Chen}{Kaplan and
  Chen}{2012}]{kaplan2012two}
Kaplan, D. and J.~Chen (2012).
\newblock A two-step {B}ayesian approach for propensity score analysis:
  Simulations and case study.
\newblock {\em Psychometrika\/}~{\em 77\/}(3), 581--609.

\bibitem[\protect\citeauthoryear{Kim}{Kim}{2011}]{kim2011parametric}
Kim, J.~K. (2011).
\newblock Parametric fractional imputation for missing data analysis.
\newblock {\em Biometrika\/}~{\em 98\/}(1), 119--132.

\bibitem[\protect\citeauthoryear{Kim and Kim}{Kim and
  Kim}{2007}]{kim2007nonresponse}
Kim, J.~K. and J.~J. Kim (2007).
\newblock Nonresponse weighting adjustment using estimated response
  probability.
\newblock {\em Canadian Journal of Statistics\/}~{\em 35\/}(4), 501--514.

\bibitem[\protect\citeauthoryear{Kim and Shao}{Kim and
  Shao}{2013}]{kim2013statistical}
Kim, J.~K. and J.~Shao (2013).
\newblock {\em Statistical Methods for Handling Incomplete Data}.
\newblock CRC Press.

\bibitem[\protect\citeauthoryear{Kim and Yu}{Kim and
  Yu}{2011}]{kim2011semiparametric}
Kim, J.~K. and C.~L. Yu (2011).
\newblock A semiparametric estimation of mean functionals with nonignorable
  missing data.
\newblock {\em Journal of the American Statistical Association\/}~{\em
  106\/}(493), 157--165.

\bibitem[\protect\citeauthoryear{Kott and Chang}{Kott and
  Chang}{2010}]{kott2010using}
Kott, P.~S. and T.~Chang (2010).
\newblock Using calibration weighting to adjust for nonignorable unit
  nonresponse.
\newblock {\em Journal of the American Statistical Association\/}~{\em
  105\/}(491), 1265--1275.

\bibitem[\protect\citeauthoryear{Little}{Little}{2012}]{little2012calibrated}
Little, R.~J. (2012).
\newblock Calibrated {B}ayes, an alternative inferential paradigm for official
  statistics.
\newblock {\em Journal of Official Statistics\/}~{\em 28\/}(3), 309.

\bibitem[\protect\citeauthoryear{McCandless, Gustafson, and Austin}{McCandless
  et~al.}{2009}]{mccandless2009bayesian}
McCandless, L.~C., P.~Gustafson, and P.~C. Austin (2009).
\newblock {B}ayesian propensity score analysis for observational data.
\newblock {\em Statistics in Medicine\/}~{\em 28\/}(1), 94--112.

\bibitem[\protect\citeauthoryear{Morgan and Winship}{Morgan and
  Winship}{2014}]{morgan2014counterfactuals}
Morgan, S.~L. and C.~Winship (2014).
\newblock {\em Counterfactuals and {C}ausal inference}.
\newblock Cambridge University Press.

\bibitem[\protect\citeauthoryear{Riddles, Kim, and Im}{Riddles
  et~al.}{2016}]{riddles2016propensity}
Riddles, M.~K., J.~K. Kim, and J.~Im (2016).
\newblock A propensity-score-adjustment method for nonignorable nonresponse.
\newblock {\em Journal of Survey Statistics and Methodology\/}~{\em 4\/}(2),
  215.

\bibitem[\protect\citeauthoryear{Robins, Rotnitzky, and Zhao}{Robins
  et~al.}{1994}]{robins1994estimation}
Robins, J.~M., A.~Rotnitzky, and L.~P. Zhao (1994).
\newblock Estimation of regression coefficients when some regressors are not
  always observed.
\newblock {\em Journal of the American statistical Association\/}~{\em
  89\/}(427), 846--866.

\bibitem[\protect\citeauthoryear{Robins, Rotnitzky, and Zhao}{Robins
  et~al.}{1995}]{robins1995analysis}
Robins, J.~M., A.~Rotnitzky, and L.~P. Zhao (1995).
\newblock Analysis of semiparametric regression models for repeated outcomes in
  the presence of missing data.
\newblock {\em Journal of the American Statistical Association\/}~{\em
  90\/}(429), 106--121.

\bibitem[\protect\citeauthoryear{Rosenbaum}{Rosenbaum}{1987}]{rosenbaum1987model}
Rosenbaum, P.~R. (1987).
\newblock Model-based direct adjustment.
\newblock {\em Journal of the American Statistical Association\/}~{\em
  82\/}(398), 387--394.

\bibitem[\protect\citeauthoryear{Rosenbaum and Rubin}{Rosenbaum and
  Rubin}{1983}]{rosenbaum1983central}
Rosenbaum, P.~R. and D.~B. Rubin (1983).
\newblock The central role of the propensity score in observational studies for
  causal effects.
\newblock {\em Biometrika\/}~{\em 70\/}(1), 41--55.

\bibitem[\protect\citeauthoryear{Rubin}{Rubin}{1976}]{rubin1976inference}
Rubin, D.~B. (1976).
\newblock Inference and missing data.
\newblock {\em Biometrika\/}~{\em 63\/}(3), 581--592.

\bibitem[\protect\citeauthoryear{Scharfstein, Rotnitzky, and
  Robins}{Scharfstein et~al.}{1999}]{scharfstein1999adjusting}
Scharfstein, D.~O., A.~Rotnitzky, and J.~M. Robins (1999).
\newblock Adjusting for nonignorable drop-out using semiparametric nonresponse
  models.
\newblock {\em Journal of the American Statistical Association\/}~{\em
  94\/}(448), 1096--1120.

\bibitem[\protect\citeauthoryear{Soubeyrand and Haon-Lasportes}{Soubeyrand and
  Haon-Lasportes}{2015}]{soubeyrand2015weak}
Soubeyrand, S. and E.~Haon-Lasportes (2015).
\newblock Weak convergence of posteriors conditional on maximum
  pseudo-likelihood estimates and implications in abc.
\newblock {\em Statistics \& Probability Letters\/}~{\em 107}, 84--92.

\bibitem[\protect\citeauthoryear{Tanner and Wong}{Tanner and
  Wong}{1987}]{tanner1987calculation}
Tanner, M.~A. and W.~H. Wong (1987).
\newblock The calculation of posterior distributions by data augmentation.
\newblock {\em Journal of the American statistical Association\/}~{\em
  82\/}(398), 528--540.

\bibitem[\protect\citeauthoryear{Van~der Vaart}{Van~der
  Vaart}{2000}]{van2000asymptotic}
Van~der Vaart, A.~W. (2000).
\newblock {\em Asymptotic {S}tatistics}, Volume~3.
\newblock Cambridge university press.

\bibitem[\protect\citeauthoryear{Wang, Shao, and Kim}{Wang
  et~al.}{2014}]{wang2014instrumental}
Wang, S., J.~Shao, and J.~K. Kim (2014).
\newblock An instrumental variable approach for identification and estimation
  with nonignorable nonresponse.
\newblock {\em Statistica Sinica\/}~{\em 24}, 1097--1116.

\bibitem[\protect\citeauthoryear{Zhou and Kim}{Zhou and
  Kim}{2012}]{zhou2012efficient}
Zhou, M. and J.~K. Kim (2012).
\newblock An efficient method of estimation for longitudinal surveys with
  monotone missing data.
\newblock {\em Biometrika\/}~{\em 99\/}(3), 631--648.

\end{thebibliography}


\end{document}